\documentclass[11pt]{article}
\usepackage{amssymb,amsmath,amsfonts}
\usepackage{graphicx}
\usepackage{graphics}
\usepackage{eepic,epsfig}

1.60cm \makeatletter \@addtoreset{equation}{section}

\makeatother

\begin{document}

\title{Casimir effect in hemisphere capped tubes}
\author{E. R. Bezerra de Mello$^{1}$\thanks{
E-mail: emello@fisica.ufpb.br} \thinspace\ and A. A. Saharian$^{1,2}$\thanks{%
E-mail: saharian@ysu.am} \\
%EndAName
\textit{$^{1}$Departamento de F\'{\i}sica, Universidade Federal da Para\'{\i}%
ba}\\
\textit{58.059-970, Caixa Postal 5.008, Jo\~{a}o Pessoa, PB, Brazil}\vspace{%
0.3cm}\\
\textit{$^2$Department of Physics, Yerevan State University,}\\
\textit{1 Alex Manoogian Street, 0025 Yerevan, Armenia}}
\maketitle

\begin{abstract}
In this paper we investigate the vacuum densities for a massive scalar field
with general curvature coupling in background of a (2+1)-dimensional
spacetime corresponding to a cylindrical tube with a hemispherical cap. A
complete set of mode functions is constructed and the positive-frequency
Wightman function is evaluated for both the cylindrical and hemispherical
subspaces. On the base of this, the vacuum expectation values of the field
squared and energy-momentum tensor are investigated. The mean field squared
and the normal stress are finite on the boundary separating two subspaces,
whereas the energy density and the parallel stress diverge as the inverse
power of the distance from the boundary. For a conformally coupled field,
the vacuum energy density is negative on the cylindrical part of the space.
On the hemisphere, it is negative near the top and positive close to the
boundary. In the case of minimal coupling the energy density on the cup is
negative. On the tube it is positive near the boundary and negative at large
distances. Though the geometries of the subspaces are different, the Casimir
pressures on the separate sides of the boundary are equal and the net
Casimir force vanishes. The results obtained may be applied to capped carbon
nanotubes described by an effective field theory in the long-wavelength
approximation.
\end{abstract}

\bigskip


\bigskip

\section{Introduction}

There are a variety of reasons for the study of field theoretical effects in
$(2+1)$-dimensional background spacetimes. In addition to be simplified
models in particle physics, $(2+1)$-dimensional theories exhibit a number of
features, such as parity violation, flavour symmetry breaking,
fractionalization of quantum numbers, that make them interesting on their
own (see Refs. \cite{Dese82}-\cite{Dunn99}). In three-dimensions,
topologically non-trivial gauge invariant terms in the action provide masses
for the gauge fields. The topological mass term introduces an infrared
cutoff in vector gauge theories providing a way for the solution of the
infrared problem without changing the ultraviolet behavior \cite{Dese82}.
Another motivation is related to the connection of three-dimensional gauge
theories to quantum chromodynamics in the high-temperature limit \cite%
{Gros81}. In addition to their fundamental interest, three-dimensional field
theoretical models appear as effective theories for various systems of
interest. In particular, in condensed matter physics they include high
temperature superconductors \cite{Fran02}, graphene \cite{Cast09}, and more
recently, topological insulators \cite{Hasa10}. Three-dimensional
topological insulators have conducting states on their boundary which are
protected by time-reversal symmetry and these states can be effectively
described in terms of $(2+1)$-dimensional Dirac fermions propagating on the
boundary.

In graphene, the low-energy excitations of the electronic subsystem are
described by an effective Dirac theory with the Fermi velocity playing the
role of speed of light (for a review see Ref. \cite{Cast09}). The
corresponding effective 3-dimensional relativistic field theory, in addition
to Dirac fermions, involves scalar and gauge fields originating from the
elastic properties and describing disorder phenomena, like the distortions
of the graphene lattice and structural defects (see, for example, Ref. \cite%
{Jack07} and references therein). For a flat graphene sheet the
fields live on $(2+1)$-dimensional Minkowski spacetime. Single-wall
cylindrical carbon nanotubes are obtained by rolling the graphene
sheet into a cylindrical shape. Though the background spacetime
remains flat, the spatial topology is changed to $R^{1}\times
S^{1}$. Another class of graphene made structures, toroidal carbon
nanotubes with the topology $S^{1}\times S^{1}$, are obtained by the
further compactification of a cylindrical tube along its axis.
One-loop quantum effects in the corresponding Dirac-like theory,
induced by non-trivial topology of graphene made cylindrical and
toroidal nanotubes, have been recently studied in Refs.
\cite{Bell09}. The finite temperature effects on the fermionic
condensate and current densities in these geometries are discussed
in \cite{Bell14T}. In reality, the cylindrical nanotubes have a
finite length. The end of a nanotube can either be open or closed by
hemispherical or conical caps \cite{Dres92}. For an open tube, the
presence of the edges imposes boundary conditions on the electron
wave function ensuring a zero flux through the edges. These boundary
conditions give rise to the Casimir effect (for reviews of the
Casimir effect and its applications in nanophysics see
\cite{Bord09,Milt02}) for the expectation values of physical
observables in effective $(2+1)$-dimensional Dirac theory. The
corresponding Casimir energy, forces and the vacuum densities are
investigated in \cite{Bell09b}.

Continuing in this line of investigations, in the present work we consider a
background geometry corresponding to a semi-infinite tube with a
hemispherical cap. In this background one has two spatial regions with
different geometrical properties separated by a circular boundary. The
geometry of one region affects the properties of the quantum vacuum in the
other region. This is a gravitational analog of the electromagnetic Casimir
effect with boundaries separating the regions having different
electromagnetic properties. Previously, we have considered examples of this
type of gravitationally induced effects for a cosmic string with a
cylindrically symmetric core with finite support \cite{Beze06} and for a
global monopole with a spherically symmetric core \cite{Beze06b}. In these
works the general results were specified for the 'ballpoint pen' model \cite%
{Hisk85}, with a constant curvature metric of the core, and for the 'flower
pot' model \cite{Alle90} with an interior Minkowskian spacetime. More
recently various types of background geometries separated by a spherical
boundary have been discussed in \cite{Milt12}. The vacuum expectation values
of the field squared and the energy-momentum tensor induced in anti-de
Sitter spacetime by a $Z_{2}$-symmetric brane with finite thickness are
evaluated in \cite{Saha07}. In the corresponding problem the boundaries
separating different spatial regions are plane symmetric.

In the present paper we evaluate the two-point function and the vacuum
expectation values (VEVs) of the field squared and energy-momentum tensor
for a scalar field in a $(2+1)$-dimensional geometry of a hemisphere capped
tube. These VEVs are among the most important characteristics of the vacuum
state. In particular, the normal stress evaluated at the boundary separating
the cylindrical and hemispherical subspace determines the Casimir force. The
organization of the paper is as follows. In the next section, the geometry
of the problem is presented and a complete set of normalized mode functions
is constructed for a scalar field with general curvature coupling. The
positive frequency Wightman function on both the cylindrical and
hemispherical subspaces are evaluated in section \ref{sec:WF}. The parts
induced by the coexisting geometries are explicitly separated. The VEV of
the field squared is investigated in section \ref{sec:phi2}. Various
asymptotics are discussed and numerical results are presented. The
corresponding analysis for the VEV of the energy-momentum tensor and for the
Casimir pressure is presented in section \ref{sec:EMT}. The main results of
the paper are summarized in section \ref{sec:Conc}. In appendix \ref%
{sec:App1} we prove the identity involving the associated Legendre functions
that is used for the evaluation of the Wightman function on the
hemispherical cap. In appendix \ref{sec:App2} expressions are derived for
the renormalized VEVs of the field squared and energy-momentum tensor on $%
S^{2}$ for general case of curvature coupling.

\section{Geometry and the mode functions}

\label{sec:Modes}

\begin{figure}[tbph]
\begin{center}
\epsfig{figure=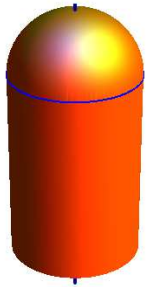,width=2.5cm,height=5.cm}
\end{center}
\caption{Cylindrical tube with a hemisphere cap.}
\label{fig1}
\end{figure}

The geometry we want to consider in this analysis is a $(2+1)$-dimensional
spacetime corresponding to a semi-cylindrical surface with topology $%
S^{1}\times R^{1}$, capped by a hemisphere (see figure \ref{fig1}). The
coordinates covering all the manifold will be denoted by $%
(x^{0}=t,x^{1},x^{2})$, where $0\leqslant x^{2}\leqslant L$, being $L$ the
length of the compact dimension in the semi-cylinder subspace. The
definition of $x^{1}$ is in according with the submanifold. The metric
tensor associated with each subspace is specified by the following line
elements:

\begin{itemize}
\item For the hemisphere we have
\begin{equation}
ds^{2}=dt^{2}-(dx^{1})^{2}-\sin ^{2}(x^{1}/a)(dx^{2})^{2},  \label{ds2in}
\end{equation}
where $0\leqslant x^{1}\leqslant \pi a/2$, with $a=L/(2\pi )$ being the
radius.

\item For the semi-cylinder we have
\begin{equation}
ds^{2}=dt^{2}-(dx^{1})^{2}-(dx^{2})^{2},  \label{ds2out}
\end{equation}%
where $\pi a/2\leqslant x^{1}<\infty $.
\end{itemize}

Note that the coordinate $x^{1}$ measures the distance from the top of the
hemisphere.

In addition to the spatial coordinates $(x^{1},x^{2})$ we shall use the
angular coordinates $(\theta ,\phi )$ defined as%
\begin{equation}
\theta =x^{1}/a,\;\phi =x^{2}/a,\;0\leqslant \phi \leqslant 2\pi .
\label{tetphi}
\end{equation}%
The angle $\phi $ is the standard azimuthal one with the range $[0,2\pi ]$;
as to $\theta $ it is the usual polar angle on the hemisphere, on which $%
0\leqslant \theta \leqslant \pi /2$ and $\theta =0$ corresponds to the top
of the hemisphere. On the semi-cylinder its range is $\pi /2\leqslant \theta
<\infty $. The circle $\theta =\pi /2$ corresponds to the boundary between
the two geometries (\ref{ds2in}) and (\ref{ds2out}) (circle in figure \ref%
{fig1}). In the coordinates $(t,\theta ,\phi )$ the metric tensor is given
by the expressions:%
\begin{equation}
g_{ik}=\left\{
\begin{array}{ll}
\mathrm{diag}(1,-a^{2},-a^{2}\sin ^{2}\theta ), & 0\leqslant \theta
\leqslant \pi /2, \\
\mathrm{diag}(1,-a^{2},-a^{2}), & \pi /2\leqslant \theta <\infty .%
\end{array}%
\right.  \label{gik}
\end{equation}%
The metric tensor and its first derivatives are continuous at $\theta =\pi
/2 $. So there is no need to introduce an additional "surface"
energy-momentum tensor located on the boundary. In the region $0\leqslant
\theta <\pi /2$, for the components of the Ricci tensor and for the
curvature scalar one has%
\begin{equation}
R_{0}^{0}=0,\;R_{i}^{k}=a^{-2}\delta _{i}^{k},\;R=2a^{-2},\;i,k=1,2.
\label{R00}
\end{equation}%
In the region $\pi /2<\theta <\infty $ the spacetime is flat. Note that the
extrinsic curvature tensor vanishes for both sides of the boundary.

Let us consider a massive quantum scalar field, $\varphi (x)$, with an
arbitrary curvature coupling parameter, $\xi $, on background of the
geometry described by the metric tensor (\ref{gik}). The corresponding
equation of motion reads
\begin{equation}
(g^{ik}\nabla _{i}\nabla _{k}+m^{2}+\xi R)\varphi (x)=0.  \label{fieldeq}
\end{equation}%
The most important special cases correspond to minimally ($\xi =0$) and
conformally ($\xi =1/8$) coupled fields. The hemisphere cap will change the
properties of the scalar vacuum on the cylindrical part of the space
compared with the case of an infinite tube. Similarly, the cylindrical
geometry will induce changes in the VEVs of physical observables on the
hemisphere, when compared to the case of the spherical geometry $S^{2}$.
These changes can be referred as geometrically induced Casimir densities. In
the problem under consideration, the only interaction of the field is with
the background geometry and all the properties of the quantum vacuum may be
deduced from two-point functions. Here we will evaluate the
positive-frequency Wightman function. The consideration of other two-point
functions is similar. For the evaluation of the Wightman function we shall
follow the direct mode-summation approach. In this approach a complete set
of normalized mode functions, $\{\varphi _{\sigma }(x),\varphi _{\sigma
}^{\ast }(x)\}$, is required. Here we have denoted by $\sigma $ the set of
quantum numbers specifying the mode functions.

In accordance with the symmetry of the problem, the mode functions can be
presented in the factorized form as,%
\begin{equation}
\varphi _{\alpha }(x)=f(\theta )e^{in\phi -i\omega t},\;n=0,\pm 1,\pm
2,\ldots  \label{phi1}
\end{equation}%
Substituting this expression into (\ref{fieldeq}) we get the equation for
the function $f(\theta )$. In the cylindrical part, the solution of the
equation is given by%
\begin{equation}
f(\theta )=C_{\mathrm{c}}\cos [p\left( \theta -\pi /2\right) -\alpha ],
\label{fout}
\end{equation}%
with constants $C_{\mathrm{c}}$ and $\alpha $, and%
\begin{equation}
p=\sqrt{\omega ^{2}a^{2}-n^{2}-m^{2}a^{2}}.  \label{p}
\end{equation}%
Here we assume that $p\geqslant 0$ . The allowance of bound states with
purely imaginary $p$ will be discussed below.

On the hemisphere, the equation for the function $f(\theta )$ reads%
\begin{equation}
\left[ \frac{1}{\sin \theta }\partial _{\theta }\left( \sin \theta \partial
_{\theta }\right) -\frac{n^{2}}{\sin ^{2}\theta }+\omega
^{2}a^{2}-m^{2}a^{2}-2\xi \right] f(\theta )=0.  \label{feqin}
\end{equation}%
The solution of this equation, regular at $\theta =0$, is given in terms of
the associated Legendre function of the first kind (for the properties of
the associated Legendre functions $P_{\nu }^{\mu }(x)$ and $Q_{\nu }^{\mu
}(x)$ on the cut $-1<x<1$ see \cite{Erde53}):%
\begin{equation}
f(\theta )=C_{\mathrm{s}}P_{\lambda -1/2}^{-|n|}(u),\;u=\cos \theta ,
\label{fin}
\end{equation}%
where%
\begin{equation}
\lambda =\sqrt{\omega ^{2}a^{2}-\omega _{m}^{2}}.  \label{lamb}
\end{equation}%
with the notation%
\begin{equation}
\omega _{m}=\sqrt{m^{2}a^{2}+2\xi -1/4}.  \label{omm}
\end{equation}%
Note that, because of the property $P_{-\lambda -1/2}^{-|n|}(u)=P_{\lambda
-1/2}^{-|n|}(u)$, the both signs of the square root in (\ref{lamb}) lead to
the same solution.

The solutions (\ref{fout}) and (\ref{fin}) contain three constants. One of
them is determined by the normalization condition and the remaining two are
determined from the matching conditions for the field and for its normal
derivative at $\theta =\pi /2$. In the problem under consideration the
derivatives of the metric tensor are continuous at the boundary and, hence,
the Ricci scalar in (\ref{fieldeq}) does not contain delta function terms.
As a consequence of this, both the field $\varphi (x)$ and its derivative $%
\partial _{\theta }\varphi (x)$ are continuous at $\theta =\pi /2$. From
these conditions we get%
\begin{equation}
e^{-2i\alpha }=\frac{pP_{\lambda -1/2}^{-|n|}+iP_{\lambda -1/2}^{-|n|\prime }%
}{pP_{\lambda -1/2}^{-|n|}-iP_{\lambda -1/2}^{-|n|\prime }},  \label{ealf}
\end{equation}%
for the phase in (\ref{fout}), and the relation%
\begin{equation}
C_{\mathrm{s}}=\frac{pC_{\mathrm{c}}}{\sqrt{(pP_{\lambda
-1/2}^{-|n|})^{2}+(P_{\lambda -1/2}^{-|n|\prime })^{2}}},  \label{Cs}
\end{equation}%
for the normalization coefficients. In these expressions we have introduced
the notations%
\begin{equation}
P_{\lambda -1/2}^{-|n|}=P_{\lambda -1/2}^{-|n|}(0),\;P_{\lambda
-1/2}^{-|n|\prime }=\partial _{u}P_{\lambda -1/2}^{-|n|}(u)|_{u=0}.
\label{Pnot}
\end{equation}%
The expressions for $P_{\lambda -1/2}^{-|n|}$ and $P_{\lambda
-1/2}^{-|n|\prime }$ in terms of the gamma function are given in \cite%
{Erde53}. By using the formula $\sin (\pi z)\Gamma (z)=\pi /\Gamma (1-z)$ we
get simpler representations:%
\begin{eqnarray}
P_{\lambda -1/2}^{-|n|} &=&\frac{2^{-|n|}\sqrt{\pi }}{\Gamma \left( \left(
|n|+\lambda +3/2\right) /2\right) \Gamma (\left( |n|-\lambda +3/2\right) /2)}%
,  \notag \\
P_{\lambda -1/2}^{-|n|\prime } &=&-\frac{2^{1-|n|}\sqrt{\pi }}{\Gamma \left(
\left( |n|+\lambda +1/2\right) /2\right) \Gamma (\left( |n|-\lambda
+1/2\right) /2)}.  \label{P0}
\end{eqnarray}%
So, the mode functions (\ref{fout}) and (\ref{fin}) are specified by the set
of quantum numbers $\sigma =(p,n)$.

The constant $C_{\mathrm{c}}$ is determined by the normalization condition%
\begin{equation}
\int_{0}^{\infty }d\theta \int_{0}^{2\pi }d\phi \sqrt{|g|}\varphi _{\sigma
}(x)\varphi _{\sigma ^{\prime }}^{\ast }(x)=\frac{\delta _{nn^{\prime }}}{%
2\omega }\delta (p-p^{\prime }).  \label{NormCond}
\end{equation}%
For $p^{\prime }=p$ the integral over $\theta $ is divergent and the
dominant contribution comes from the large values of $\theta $. In this
case, in order to determine $C_{\mathrm{c}}$, it is sufficient to consider
the part in the integral over the cylindrical geometry, $\theta >\pi /2$. In
this way we find%
\begin{equation}
|C_{\mathrm{c}}|^{2}=\frac{1}{2\pi ^{2}a^{2}\omega }.  \label{Cc}
\end{equation}%
Now, the mode functions with the continuous energy spectrum are written as%
\begin{eqnarray}
\varphi _{\sigma }(x) &=&C_{\mathrm{s}}P_{\lambda -1/2}^{-|n|}(\cos \theta
)e^{in\phi -i\omega t},\;0\leqslant \theta \leqslant \pi /2,  \label{phi2S2}
\\
\varphi _{\sigma }(x) &=&C_{\mathrm{c}}\cos [p\left( \theta -\pi /2\right)
-\alpha ]e^{in\phi -i\omega t},\;\pi /2\leqslant \theta <\infty ,
\label{phi2}
\end{eqnarray}%
where $0\leqslant p<\infty $ and $C_{\mathrm{s}}$ is given by (\ref{Cs}).

We can have also bound states for which $p$ is pure imaginary, $p=i\eta $
with $\eta >0$. From the stability of the vacuum one has $\omega
^{2}\geqslant 0$ and from (\ref{p}) we obtain $\eta \leqslant ma$. For the
bound states the solution in the part of the cylindrical geometry has the
form%
\begin{equation}
\varphi _{\mathrm{b}\sigma }(x)=C_{\mathrm{bc}}e^{-\eta \theta +in\phi
-i\omega t}.  \label{phib}
\end{equation}%
On the cup the corresponding solution is given by (\ref{phi2S2}) with $C_{%
\mathrm{s}}$ replaced by a new constant $C_{\mathrm{bs}}$. From the
continuity of the mode functions and their first derivatives at $\theta =\pi
/2$ one gets the equation%
\begin{equation}
P_{\lambda -1/2}^{-|n|\prime }-\eta P_{\lambda -1/2}^{-|n|}=0,  \label{bs}
\end{equation}%
where%
\begin{equation}
\lambda =\sqrt{n^{2}-\eta ^{2}-2\xi +1/4}.  \label{lambs}
\end{equation}%
Now, by using the relation
\begin{equation}
P_{\lambda -1/2}^{-|n|\prime }=\frac{-1/P_{\lambda -1/2}^{-|n|}}{\Gamma
(|n|+\lambda +1/2)\Gamma (|n|-\lambda +1/2)},  \label{RelP0}
\end{equation}%
which directly follows from (\ref{P0}), the equation for the bound states is
rewritten as%
\begin{equation}
\frac{1}{\Gamma (|n|+\lambda +1/2)\Gamma (|n|-\lambda +1/2)}+\eta
(P_{\lambda -1/2}^{-|n|})^{2}=0.  \label{bs2}
\end{equation}%
For purely imaginary $\lambda $ the lhs of the above expression is always
positive and we conclude that in this case there are no bound states. The
same is true in the case $0\leqslant \lambda \leqslant |n|+1/2$. Hence, the
bound states may be present in the case $\lambda >|n|+1/2$ only. Combining
this with (\ref{lambs}), we conclude that the bound states may exist under
the condition $\xi <-|n|/2$ only. In particular, there are no bound states
for minimally and conformally coupled fields. In what follows we assume that
$\xi \geqslant 0$ and, hence, the bound states are absent.

\section{Two-point function}

\label{sec:WF}

Having the complete set of normalized mode functions (\ref{phi2S2}) and (\ref%
{phi2}), we can evaluate any of two-point functions for a scalar field. In
particular, for the positive frequency Wightman function one has the
mode-sum formula%
\begin{equation}
W(x,x^{\prime })=\sum_{n=-\infty }^{+\infty }\int_{0}^{\infty }dp\,\varphi
_{\sigma }(x)\varphi _{\sigma }^{\ast }(x^{\prime }).  \label{WF}
\end{equation}%
First we consider the region $\pi /2<\theta <\infty $ corresponding to the
part of the space with cylindrical geometry.

\subsection{Cylindrical geometry}

Substituting the corresponding mode functions from (\ref{phi2}) into (\ref%
{WF}) and using the expression for $\alpha $, given by (\ref{ealf}), the
Wightman function is presented in the form%
\begin{eqnarray}
W(x,x^{\prime }) &=&W_{0}(x,x^{\prime })+\frac{1}{L^{2}}\sideset{}{'}{\sum}%
_{n=0}^{\infty }\cos (n\Delta \phi )\int_{0}^{\infty }dp  \notag \\
&&\times \frac{e^{-i\omega \Delta t}}{\omega }\sum_{s=\pm 1}e^{sip(\theta
+\theta ^{\prime }-\pi )}\frac{pP_{\lambda -1/2}^{-n}+siP_{\lambda
-1/2}^{-n\prime }}{pP_{\lambda -1/2}^{-n}-siP_{\lambda -1/2}^{-n\prime }},
\label{WFcyl}
\end{eqnarray}%
where $\Delta \phi =\phi -\phi ^{\prime }$, $\Delta t=t-t^{\prime }$, $%
\omega $ is expressed in terms of $p$ by using (\ref{p}), and the prime on
the sign of the sum means that the term $n=0$ should be taken with the
coefficient 1/2. In (\ref{WFcyl}),%
\begin{equation}
W_{0}(x,x^{\prime })=\frac{2}{L^{2}}\sideset{}{'}{\sum}_{n=0}^{\infty }\cos
(n\Delta \phi )\int_{0}^{\infty }dp\,\cos (p\Delta \theta )\frac{e^{-i\omega
\Delta t}}{\omega },  \label{WF0}
\end{equation}%
with $\Delta \theta =\theta -\theta ^{\prime }$, is the Wightman function
for an infinite tube described by the line element (\ref{ds2out}) for $%
-\infty <\theta <+\infty $. By using the Abel-Plana summation formula (see,
for instance, \cite{Bord09,Saha07Rev}) for the series over $n$ in (\ref{WF0}%
), this function can be presented in the form%
\begin{equation}
W_{0}(x,x^{\prime })=\frac{1}{4\pi a}\sum_{l=-\infty }^{\infty }\frac{e^{-ma%
\sqrt{(\Delta \theta )^{2}+\left( \Delta \phi +2\pi l\right) ^{2}-(\Delta
t/a)^{2}}}}{\sqrt{(\Delta \theta )^{2}+\left( \Delta \phi +2\pi l\right)
^{2}-(\Delta t/a)^{2}}}.  \label{WF02}
\end{equation}%
In this representation, the $l=0$ term is the Wightman function for
Minkowski spacetime with spatial topology $R^{2}$. The formula (\ref{WF02})
presents the Wightman function for an infinite tube as an image sum of the
Minkowskian functions.

The second term in the rhs of (\ref{WFcyl}) is induced by the hemisphere
cap. For the further transformation of this part, under the condition $%
\theta +\theta ^{\prime }>\pi +\Delta t/a$, we rotate the integration
contour over $p$ by the angle $\pi /2$ for the term with $s=+1$ and by the
angle $-\pi /2$ for the term with $s=-1$. The integrals over the intervals $%
(0,ima)$ and $(0,-ima)$ cancel out and, after some transformations, one finds%
\begin{eqnarray}
W(x,x^{\prime }) &=&W_{0}(x,x^{\prime })+\frac{1}{2\pi ^{2}a}%
\sideset{}{'}{\sum}_{n=0}^{\infty }\cos (n\Delta \phi )\int_{ma}^{\infty
}dy\,y  \notag \\
&&\times \frac{\cosh (\sqrt{y^{2}-m^{2}a^{2}}\Delta t/a)}{\sqrt{%
y^{2}-m^{2}a^{2}}}\frac{e^{-\sqrt{y^{2}+n^{2}}(\theta +\theta ^{\prime }-\pi
)}}{\sqrt{y^{2}+n^{2}}}f_{n}(y).  \label{WFcyl2}
\end{eqnarray}%
Here we have introduced the notation%
\begin{equation}
f_{n}(y)=\frac{\sqrt{y^{2}+n^{2}}P_{iz(y)-1/2}^{-n}+P_{iz(y)-1/2}^{-n\prime }%
}{\sqrt{y^{2}+n^{2}}P_{iz(y)-1/2}^{-n}-P_{iz(y)-1/2}^{-n\prime }},
\label{fn}
\end{equation}%
with%
\begin{equation}
z(y)=\sqrt{y^{2}+2\xi -1/4}.  \label{zy}
\end{equation}%
Note that the function $z(y)$ can be either real or purely imaginary. In
both cases the function $P_{iz(y)-1/2}^{-n}$ is real. For $\xi \geqslant 0$
the function in the denominator of (\ref{fn}) is positive. Unlike to the
oscillating integrand in (\ref{WFcyl}), the integrand in (\ref{WFcyl2}),
under the condition mentioned above, is exponentially decreasing near the
upper limit of the integration and the representation (\ref{WFcyl2}) is well
adapted for the evaluation of the VEVs in the coincidence limit.

\subsection{Hemisphere cap}

In this subsection we consider the two-point function on the hemispherical
cap (for the zeta function and heat kernel coefficients on Riemann caps see
\cite{Flac11}). From the mode-sum formula (\ref{WF}) with the mode functions
from (\ref{phi2S2}), for the Wightman function on the hemisphere cap, one has%
\begin{equation}
W(x,x^{\prime })=\frac{4}{L^{2}}\sideset{}{'}{\sum}_{n=0}^{\infty }\cos
(n\Delta \phi )\int_{0}^{\infty }dp\,p^{2}\frac{P_{\lambda -1/2}^{-n}(\cos
\theta )P_{\lambda -1/2}^{-n}(\cos \theta ^{\prime })}{(pP_{\lambda
-1/2}^{-n})^{2}+(P_{\lambda -1/2}^{-n\prime })^{2}}\frac{e^{-i\omega \Delta
t}}{\omega },  \label{WFcap}
\end{equation}%
with $0\leqslant \theta ,\theta ^{\prime }<\pi /2$ and $\lambda $ is given
by (\ref{lamb}). Our interest in this paper is the effects on the sphere,
induced by the cylindrical geometry, in the region $\theta >\pi /2$. In
order to explicitly extract from (\ref{WFcap}) the part induced by this
geometry, we note that the denominator in the integrand of (\ref{WFcap}) is
equal to $\bar{P}_{\lambda -1/2}^{+,-|n|}(0)\bar{P}_{\lambda
-1/2}^{-,-|n|}(0)$, where the notations with bars are defined by the
relation (\ref{BarNot}) in the appendix \ref{sec:App1}. By using the
identity (\ref{Ident2}) with $\theta =\pi /2$, the function (\ref{WFcap}) is
expressed as%
\begin{eqnarray}
W(x,x^{\prime }) &=&\frac{i}{L^{2}}\sideset{}{'}{\sum}_{n=0}^{\infty
}(-1)^{n}\cos (n\Delta \phi )\int_{0}^{\infty }dp\,p\frac{e^{-i\omega \Delta
t}}{\omega }  \notag \\
&&\times P_{\lambda -1/2}^{-n}(\cos \theta )P_{\lambda -1/2}^{n}(\cos \theta
^{\prime })\sum_{j=+,-}j\frac{\bar{S}_{\lambda -1/2}^{j,-n}(0)}{\bar{P}%
_{\lambda -1/2}^{j,-n}(0)},  \label{WFcap2}
\end{eqnarray}%
where the functions $\bar{S}_{\lambda -1/2}^{j,-n}(x)$ are defined by the
expressions (\ref{S}) and (\ref{Sbar}) in appendix \ref{sec:App1} and we
have used the relation \cite{Erde53}%
\begin{equation}
P_{\lambda -1/2}^{-n}(u)=(-1)^{n}\frac{\Gamma \left( \lambda -n+1/2\right) }{%
\Gamma \left( \lambda +n+1/2\right) }P_{\lambda -1/2}^{n}(u).  \label{Prel}
\end{equation}%
Note that the integrand in (\ref{WFcap2}) is an even function of $\lambda $.

The expressions for $S_{\lambda -1/2}^{-n}(0)$ and $S_{\lambda
-1/2}^{-n\prime }(0)$, entering in (\ref{WFcap2}), in terms of the gamma
function are found by using the corresponding expressions for $Q_{\lambda
-1/2}^{-n}(0)$ and $Q_{\lambda -1/2}^{-n\prime }(0)$ from \cite{Erde53}. The
latter can be written as%
\begin{eqnarray}
Q_{\lambda -1/2}^{-n}(0) &=&-\frac{\pi }{2}\tan \left[ \pi (\lambda -n-1/2)/2%
\right] P_{\lambda -1/2}^{-n},  \notag \\
Q_{\lambda -1/2}^{-n\prime }(0) &=&\frac{\pi }{2}\cot \left[ \pi (\lambda
-n-1/2)/2\right] P_{\lambda -1/2}^{-n\prime },  \label{Q0}
\end{eqnarray}%
with $P_{\lambda -1/2}^{-n}$ and $P_{\lambda -1/2}^{-n\prime }$ given in (%
\ref{P0}). By using these relations, we can see that%
\begin{eqnarray}
S_{\lambda -1/2}^{-n}(0) &=&\pi (-1)^{n}\frac{P_{\lambda -1/2}^{-n}}{\cos
(\pi \lambda )},  \notag \\
S_{\lambda -1/2}^{-n\prime }(0) &=&-\pi (-1)^{n}\frac{P_{\lambda
-1/2}^{-n\prime }}{\cos (\pi \lambda )}.  \label{S0}
\end{eqnarray}

By taking into account that in the complex plane $p$ and under the condition
$\pi -\theta -\theta ^{\prime }-\Delta t/a>0$, the integrand of (\ref{WFcap2}%
) exponentially decreases for $|\mathrm{Im}\,p|\rightarrow \infty $, we
rotate the integration contour over $p$ by the angle $\pi /2$ ($-\pi /2$)
for the term with $j=+$ ($j=-$). Note that, though the integrand in (\ref%
{WFcap2}) has no poles on the real axis, this is not the case for separate
terms in the sum over $j$. By taking into account the relation (\ref{S2}),
we see that the separate terms have poles at the zeros of the function $\cos
(\lambda \pi )$. In the integral over $p$ we will shift the integration
contour by a small amount $\epsilon >0$ assuming that the integration goes
over the half-line $[0-i\epsilon ,\infty -i\epsilon )$. Now, in the rotation
for the $j=+$ term, we note that the integrand has poles at the zeros of the
function $\cos (\lambda \pi )$ and the corresponding residues should be
included. After the rotation of the contours, the integrals over the regions
$(0,i\sqrt{n^{2}+m^{2}a^{2}})$ and $(0,-i\sqrt{n^{2}+m^{2}a^{2}})$ cancel
out. In this way, by using the relations (\ref{S0}), from (\ref{WFcap2}) one
gets%
\begin{eqnarray}
W(x,x^{\prime }) &=&W_{S^{2}}(x,x^{\prime })-\frac{1}{L}\sideset{}{'}{\sum}%
_{n=0}^{\infty }\cos (n\Delta \phi )\int_{ma}^{\infty }dy\,\frac{yf_{n}(y)}{%
\cosh (\pi z(y))}  \notag \\
&&\times P_{iz(y)-1/2}^{-n}(\cos \theta )P_{iz(y)-1/2}^{n}(\cos \theta
^{\prime })\frac{\cosh (\sqrt{y^{2}-m^{2}a^{2}}\Delta t/a)}{\sqrt{%
y^{2}-m^{2}a^{2}}},  \label{WFcap3}
\end{eqnarray}%
where the functions $f_{n}(y)$ and $z(y)$ were already defined by the
relations (\ref{fn}) and (\ref{zy}). The first term in the rhs of (\ref%
{WFcap3}) comes from the residues at the zeros of the function $\cos
(\lambda \pi )$ and coincides with the Wightman function for a scalar field
on the sphere $S^{2}$ (see the appendix \ref{sec:App2}). After the summation
over $n$ by using the formula \cite{Erde53}%
\begin{equation}
\sum_{n=-l}^{+l}(-1)^{|n|}P_{l}^{-|n|}(\cos \theta )P_{l}^{|n|}(\cos \theta
^{\prime })e^{in\Delta \phi }=P_{l}(\cos \gamma ),  \label{SumForm}
\end{equation}%
this function is expressed as%
\begin{equation}
W_{S^{2}}(x,x^{\prime })=\frac{a^{-2}}{8\pi }\sum_{l=0}^{\infty }\frac{2l+1}{%
\omega _{l}}e^{-i\omega _{l}\Delta t}P_{l}(\cos \gamma ).  \label{WS2}
\end{equation}%
In this formula, $P_{l}(z)$ is the Legendre polynomial and
\begin{eqnarray}
\omega _{l} &=&a^{-1}\sqrt{l(l+1)+m^{2}a^{2}+2\xi },  \notag \\
\cos \gamma &=&\cos \theta \cos \theta ^{\prime }+\sin \theta \sin \theta
^{\prime }\cos (\Delta \phi ).  \label{oml}
\end{eqnarray}

When compared with (\ref{WFcap2}), the representation (\ref{WFcap3}) has two
important advantages. First of all, the part describing the effects induced
by the cylindrical tube is explicitly extracted. In this way, for points
away from the boundary $\theta =\pi /2$, the renormalization of the VEVs in
the coincidence limit is reduced to the one for the geometry $S^{2}$. And
second, under the condition $\pi -\theta -\theta ^{\prime }-\Delta t/a>0$,
the integrand in the tube-induced part of (\ref{WFcap3}) exponentially
decreases near the upper limit of the integration. This is important in the
numerical evaluation of the local VEVs in the coincidence limit.

\section{Mean field squared}

\label{sec:phi2}

Formally the VEV of the field squared is obtained from the Wightman function
by taking the coincidence limit of the arguments. However, this procedure
results in a divergent quantity and some renormalization procedure is needed
to provide a well defined finite value. We shall consider the cylindrical
and hemisphere parts separately.

Let us start with the part of the geometry corresponding to the cylindrical
tube. In this subspace the spacetime is flat and the renormalization is
reduced to the subtraction from the VEVs the corresponding VEVs in $(2+1)$%
-dimensional Minkowski spacetime with trivial topology. The latter is given
by the $l=0$ term in (\ref{WF02}). Omitting this term and using the
expression (\ref{WFcyl2}) for the Wightman function, the renormalized VEV of
the field squared is expressed in the form%
\begin{eqnarray}
\langle \varphi ^{2}\rangle &=&\langle \varphi ^{2}\rangle _{0}+\frac{1}{\pi
L}\sideset{}{'}{\sum}_{n=0}^{\infty }\int_{ma}^{\infty }dy\,  \notag \\
&&\times \frac{yf_{n}(y)}{\sqrt{y^{2}-m^{2}a^{2}}}\frac{e^{-\sqrt{y^{2}+n^{2}%
}(2\theta -\pi )}}{\sqrt{y^{2}+n^{2}}},  \label{phi2cyl}
\end{eqnarray}%
for $\theta >\pi /2$. Here%
\begin{equation}
\langle \varphi ^{2}\rangle _{0}=-\frac{1}{2\pi L}\ln (1-e^{-Lm}),
\label{phi2cyl0}
\end{equation}%
is the VEV for an infinite tube and the second term in the rhs is induced by
the cap. Note that for a massless field the VEV contains infrared
divergences.

For large values of $n$ and $y$, by using the asymptotic formula for the
gamma function for large values of the argument, we can see that to the
leading order
\begin{equation}
f_{n}(y)\approx \frac{(y^{2}+n^{2})\left( 1-4\xi \right) -n^{2}}{%
8(y^{2}+n^{2})^{2}}.  \label{fnas}
\end{equation}%
From here it follows that the cap-induced part in (\ref{phi2cyl}) is finite
for all values of $\pi /2\leqslant \theta <\infty $, including the points on
the boundary $\theta =\pi /2$. At large distances from the tube edge, $%
a(\theta -\pi /2)\gg 1/m$, the dominant contribution to the integral in (\ref%
{phi2cyl}) comes from the region near the lower limit and to the leading
order we get%
\begin{equation}
\langle \varphi ^{2}\rangle \approx \langle \varphi ^{2}\rangle _{0}+\frac{%
f_{0}(ma)e^{-ma(2\theta -\pi )}}{4L\sqrt{\pi ma(\theta -\pi /2)}}.
\label{phi2cyllarge}
\end{equation}%
In this limit the cap-induced contribution is exponentially small. The
expression (\ref{phi2cyllarge}) describes also the behavior of the
cap-induced part in the limit of large values for the tube radius with fixed
value of $\theta -\pi /2$. Note that in this limit one has $\langle \varphi
^{2}\rangle _{0}\approx e^{-Lm}/(2\pi L)$ and for $\theta <3\pi /2$ the
total VEV\ is dominated by the cap-induced contribution.

It will be interesting to compare the expressions for the hemisphere capped
tube with the corresponding results for a semi-infinite (si) tube with
Dirichlet or Neumann boundary conditions at the edge $\theta =\pi /2$. The
Wightman function and the VEV of the field squared in the latter geometry
are obtained from (\ref{WFcyl2}) and (\ref{phi2cyl}) by the replacement $%
f_{n}(y)\rightarrow \mp 1$, where the upper and lower signs correspond to
Dirichlet and Neumann boundary conditions, respectively. With this
replacement, after the integration over $y$, we get%
\begin{equation}
\langle \varphi ^{2}\rangle _{\mathrm{si}}=\langle \varphi ^{2}\rangle
_{0}\mp \frac{1}{\pi L}\sideset{}{'}{\sum}_{n=0}^{\infty }K_{0}((2\theta
-\pi )\sqrt{n^{2}+m^{2}a^{2}}),  \label{phi2si}
\end{equation}%
where $K_{\nu }(x)$ is the MacDonald function. The second term in the rhs of
(\ref{phi2si}) is induced by the edge of the tube. For Neumann boundary
condition the VEV (\ref{phi2si}) is positive, whereas for Dirichlet boundary
condition it is negative near the edge and positive at large distances. For
points close to the boundary, the contribution of large $n$ dominates in (%
\ref{phi2si}) and, to the leading order, the summation can be replaced by
the integration. In this way, we see that for a semi-infinite tube with
Dirichlet or Neumann boundary conditions the VEV of the field squared
diverges on the boundary as $1/(2\theta -\pi )$.

In the part of the geometry corresponding to the hemisphere cap we use the
expression (\ref{WFcap3}) for the Wightman function. In the coincidence
limit this gives%
\begin{equation}
\langle \varphi ^{2}\rangle =\langle \varphi ^{2}\rangle _{S^{2}}-\frac{1}{L}%
\sideset{}{'}{\sum}_{n=0}^{\infty }\int_{ma}^{\infty }dy\,\frac{yg_{n}(y)}{%
\sqrt{y^{2}-m^{2}a^{2}}}\frac{[P_{iz(y)-1/2}^{-n}(\cos \theta )]^{2}}{\cosh
[\pi z(y)]},  \label{phi2cap}
\end{equation}%
where $\theta <\pi /2$ and we have introduced the notation%
\begin{eqnarray}
g_{n}(y) &=&(-1)^{n}\frac{\Gamma (iz(y)+n+1/2)}{\Gamma (iz(y)-n+1/2)}f_{n}(y)
\notag \\
&=&\frac{\sqrt{y^{2}+n^{2}}P_{iz(y)-1/2}^{n}+P_{iz(y)-1/2}^{n\prime }}{\sqrt{%
y^{2}+n^{2}}P_{iz(y)-1/2}^{-n}-P_{iz(y)-1/2}^{-n\prime }}.  \label{gny}
\end{eqnarray}%
The renormalization is reduced to that for the first term in the rhs,
corresponding to the geometry $S^{2}$. The expression for the renormalized $%
\langle \varphi ^{2}\rangle _{S^{2}}$ is derived in the appendix \ref%
{sec:App2}. The second term in the rhs of (\ref{phi2cap}) is induced by the
cylindrical tube. For large values of $y$ the function $P_{iz(y)-1/2}^{-n}(%
\cos \theta )$ behaves as $e^{y\theta }$, and the integrand in (\ref{phi2cap}%
) decays as $e^{-y(2\theta -\pi )}$ for $\theta <\pi /2$. By taking into
account the asymptotic expression (\ref{fnas}), we see that the tube-induced
part is finite at the boundary, $\theta =\pi /2$.

\begin{figure}[tbph]
\begin{center}
\begin{tabular}{cc}
\epsfig{figure=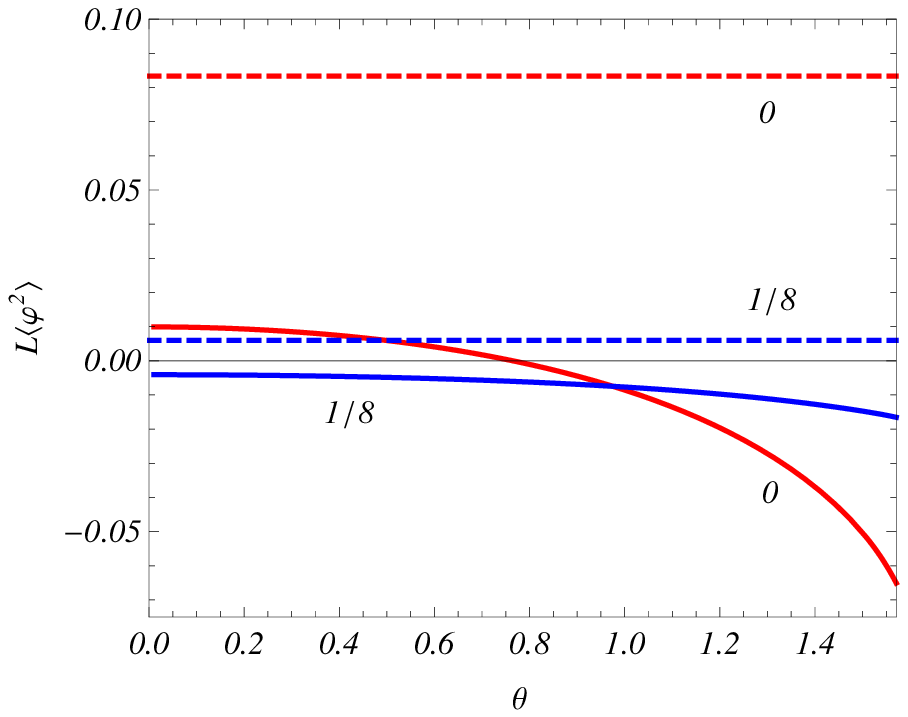,width=7.cm,height=5.5cm} & \quad %
\epsfig{figure=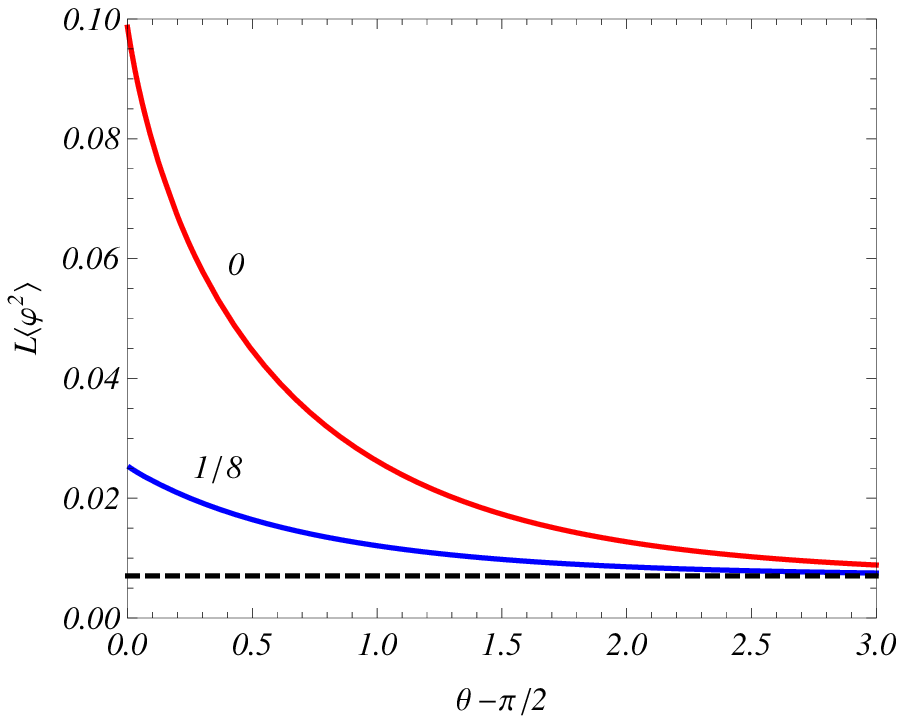,width=7.cm,height=5.5cm}%
\end{tabular}%
\end{center}
\caption{Renormalized VEV of the field squared on the hemisphere cap (left
panel) and on the cylindrical tube (right panel) for minimally and
conformally coupled scalar fields. The numbers near the curves are the
values of the curvature coupling parameter $\protect\xi $ and the graphs are
plotted for $ma=1/2$. The dashed curves display the renormalized VEVs on $%
S^{2}$ (left panel) and on an infinite tube (right panel).}
\label{fig2}
\end{figure}

In figure \ref{fig2} we have plotted the renormalized mean field squared as
a function of $\theta $ in hemispherical (left panel) and cylindrical (right
panel) regions for minimally and conformally coupled field (the numbers near
the curves are the values of the curvature coupling parameter $\xi $). The
graphs are plotted for $ma=1/2$. The dashed curves present the renormalized
VEV of the field squared on $S^{2}$ (left panel) and on an infinite tube
(right panel). In the cylindrical part, the VEV $\langle \varphi ^{2}\rangle
_{0}$ does not depend on the curvature coupling. On the hemispherical cap,
the tube-induced contribution in the VEV of the field squared is negative
for both the minimal and conformal couplings, whereas on the tube the
cap-induced parts are positive.

As is seen from fig \ref{fig2}, though the Wightman function is continuous
at the boundary, the renormalized VEV of the field squared is not
continuous. The reason for this is that in the renormalization procedure for
separate regions of tube and hemisphere we have subtracted different terms.
In the tube part of the geometry the subtracted term coincides with the
corresponding VEV in Minkowski spacetime, whereas on the hemisphere, in
addition to the Minkowskian term, we have also subtracted a finite
renormalization term $(1/6-\xi )/(4\pi ma^{2})$. We have numerically checked
that the VEV evaluated by the minimal subtraction scheme (on the hemisphere
the Minkowskian part is subtracted only), $\langle \varphi ^{2}\rangle _{%
\mathrm{min}}=\langle \varphi ^{2}\rangle +(1/6-\xi )/(4\pi ma^{2})$, is
continuous on the boundary.

\section{Energy-momentum tensor and the Casimir force}

\label{sec:EMT}

In this section, we shall study the vacuum energy-momentum tensor and the
Casimir force for the system under consideration. Given the Wightman
function and the VEV of the field squared, the VEV of the energy-momentum
tensor is evaluated by using the formula
\begin{equation}
\langle T_{ik}\rangle =\lim_{x^{\prime }\rightarrow x}\partial _{i^{\prime
}}\partial _{k}W(x,x^{\prime })+\left[ \left( \xi -1/4\right) g_{ik}\nabla
_{p}\nabla ^{p}-\xi \nabla _{i}\nabla _{k}-\xi R_{ik}\right] \langle \varphi
^{2}\rangle ,  \label{EMT}
\end{equation}%
where $R_{ik}$ is the Ricci tensor. In the rhs of this formula we have used
the expression for the energy-momentum tensor of a scalar filed which
differs from the standard expression \cite{Birr82} by the term which does
not contribute to the VEVs (see \cite{Saha04}).

\subsection{Casimir densities on the tube}

On the tube, $\pi /2<\theta <\infty $, the off-diagonal components of the
vacuum energy-momentum tensor vanish and the diagonal components are
decomposed as (with no summation over $l$)%
\begin{equation}
\langle T_{l}^{l}\rangle =\langle T_{l}^{l}\rangle _{0}-\frac{a^{-3}}{2\pi
^{2}}\sideset{}{'}{\sum}_{n=0}^{\infty }\int_{ma}^{\infty }dy\,\frac{%
yf_{n}(y)}{\sqrt{y^{2}-m^{2}a^{2}}}\frac{e^{-\sqrt{y^{2}+n^{2}}(2\theta -\pi
)}}{\sqrt{y^{2}+n^{2}}}F_{n}^{(l)}(y),  \label{Tllcyl}
\end{equation}%
where the first term in the rhs corresponds to the geometry of an infinite
tube and the second term is induced by the hemisphere cap. For the functions
in the latter one has $F_{n}^{(1)}(y)=0$ and
\begin{eqnarray}
F_{n}^{(0)}(y) &=&4\xi \left( y^{2}+n^{2}\right) -n^{2}-m^{2}a^{2},  \notag
\\
F_{n}^{(2)}(y) &=&\left( 4\xi -1\right) \left( y^{2}+n^{2}\right) +n^{2}.
\label{Fn2}
\end{eqnarray}%
Hence, the cap-induced contribution to the vacuum stress normal to the
boundary vanishes on the tube. This result could be obtained by general
arguments, based on the covariant conservation equation $\nabla _{k}\langle
T_{i}^{k}\rangle =0$. From the symmetry of the problem it follows that the
cap-induced part depend only on the coordinate $\theta $ and this equation
is reduced to a single constraint $\partial _{\theta }\langle
T_{1}^{1}\rangle _{\mathrm{c}}=0$, where $\langle T_{1}^{1}\rangle _{\mathrm{%
c}}$ is the cap-induced part (second term in the rhs of (\ref{Tllcyl})).
Now, from the condition $\langle T_{1}^{1}\rangle _{\mathrm{c}}\rightarrow 0$
for $\theta \rightarrow \infty $ it follows that $\langle T_{1}^{1}\rangle _{%
\mathrm{c}}=0$.

The part of the VEV corresponding to an infinite tube does not depend on the
curvature coupling parameter and is expressed as (no summation over $l$):%
\begin{equation}
\langle T_{l}^{l}\rangle _{0}=\frac{F^{(l)}(mL)}{2\pi L^{3}},  \label{Tll0}
\end{equation}%
where we have introduced the notations%
\begin{eqnarray}
F^{(0)}(x) &=&F^{(1)}(x)=-\mathrm{Li}_{3}(e^{-x})-x\,\mathrm{Li}_{2}(e^{-x}),
\notag \\
F^{(2)}(x) &=&2\mathrm{Li}_{3}(e^{-x})+2x\,\mathrm{Li}_{2}(e^{-x})-x^{2}\ln
(1-e^{-x}),  \label{F2}
\end{eqnarray}%
with $\mathrm{Li}_{j}(x)=\sum_{n=1}^{\infty }x^{n}/n^{j}$ being the
polylogarithm function. For a massless field one has (for the Casimir effect
in topologically nontrivial three-dimensional spacetimes see \cite{Bord09})%
\begin{equation}
\langle T_{0}^{0}\rangle _{0}=\langle T_{1}^{1}\rangle _{0}=-\frac{1}{2}%
\langle T_{2}^{2}\rangle _{0}=-\frac{\zeta (3)}{2\pi L^{3}},  \label{tll0m0}
\end{equation}%
where $\zeta (s)$ is the Riemann zeta function. For a massive field and for
large values of the tube radius, $mL\gg 1$, to the leading order we get%
\begin{equation}
\langle T_{0}^{0}\rangle _{0}\approx \frac{-1}{mL}\langle T_{2}^{2}\rangle
_{0}\approx -\frac{me^{-Lm}}{2\pi L^{2}}\,.  \label{T00tube}
\end{equation}%
As an additional check we can see that both the cylindrical and cap-induced
parts obey the trace relation
\begin{equation}
\langle T_{l}^{l}\rangle =2\left( \xi -1/8\right) \nabla _{l}\nabla
^{l}\langle \varphi ^{2}\rangle +m^{2}\langle \varphi ^{2}\rangle .
\label{TraceRel}
\end{equation}

Unlike to the VEV of the field squared, the energy density and the stress $%
\langle T_{2}^{2}\rangle $ in (\ref{Tllcyl}) diverge on the boundary $\theta
=\pi /2$. These divergences come from the cap-induced part. In order to find
the leading terms in the corresponding asymptotic expansions, we note that
for points near the boundary the dominant contribution to the cap-induced
part in (\ref{Tllcyl}) comes from large values of $n$ and $y$. By taking
into account (\ref{fnas}), to the leading order we get (no summation over $l$%
)%
\begin{equation}
\langle T_{l}^{l}\rangle \approx \frac{2\xi \left( 4\xi -1\right) +(l+1)/16}{%
16\pi (2\theta -\pi )a^{3}},  \label{NearB}
\end{equation}%
with $l=0,2$. For both minimally and conformally coupled fields the parallel
stress is positive near the boundary, whereas the energy density is positive
for a minimally coupled field and negative for the conformal coupling.

At large distances from the edge of the tube and for a massive field, $%
a(\theta -\pi /2)\gg 1/m$, the dominant contribution to the cap-induced part
in (\ref{Tllcyl}) comes from the term $n=0$ and from the region of the
integration near the lower limit. To the leading order we find (no summation
over $l=0,2$)%
\begin{equation}
\langle T_{l}^{l}\rangle \approx \langle T_{l}^{l}\rangle _{0}-\frac{4\xi -1%
}{4\pi ^{2}a^{3}}\frac{(ma)^{3/2}f_{0}(ma)}{\sqrt{2(2\theta /\pi -1)}}%
e^{-ma(2\theta -\pi )},  \label{TllLarge}
\end{equation}%
and the VEV is exponentially small. For a massless field, at large
distances, $\theta -\pi /2\gg 1$, one has a power-law decay (no summation
over $l=0,2$):%
\begin{equation}
\langle T_{l}^{l}\rangle \approx \langle T_{l}^{l}\rangle _{0}+\frac{a^{-3}}{%
4\pi ^{2}}\frac{4\xi -\delta _{l}^{2}}{(2\theta -\pi )^{2}}.
\label{Tllm0Large}
\end{equation}

For a semi-infinite tube with Dirichlet or Neumann boundary conditions on
its edge, $\theta =\pi /2$, the VEV of the energy-momentum tensor is
obtained from (\ref{Tllcyl}) with the replacement $f_{n}(y)\rightarrow \mp 1$%
, where the upper and lower signs are for Dirichlet and Neumann conditions,
respectively. In this case, the integral over $y$ is expressed in terms of
the MacDonald functions $K_{0}(x)$ and $K_{1}(x)$ with the same argument as
in (\ref{phi2si}). Near the boundary and for non-conformally coupled fields
the leading terms in the energy density and the parallel stress behave like $%
1/(2\theta -\pi )^{3}$. As we see, the divergences here are stronger than
those in the problem under consideration.

\begin{figure}[tbph]
\begin{center}
\begin{tabular}{cc}
\epsfig{figure=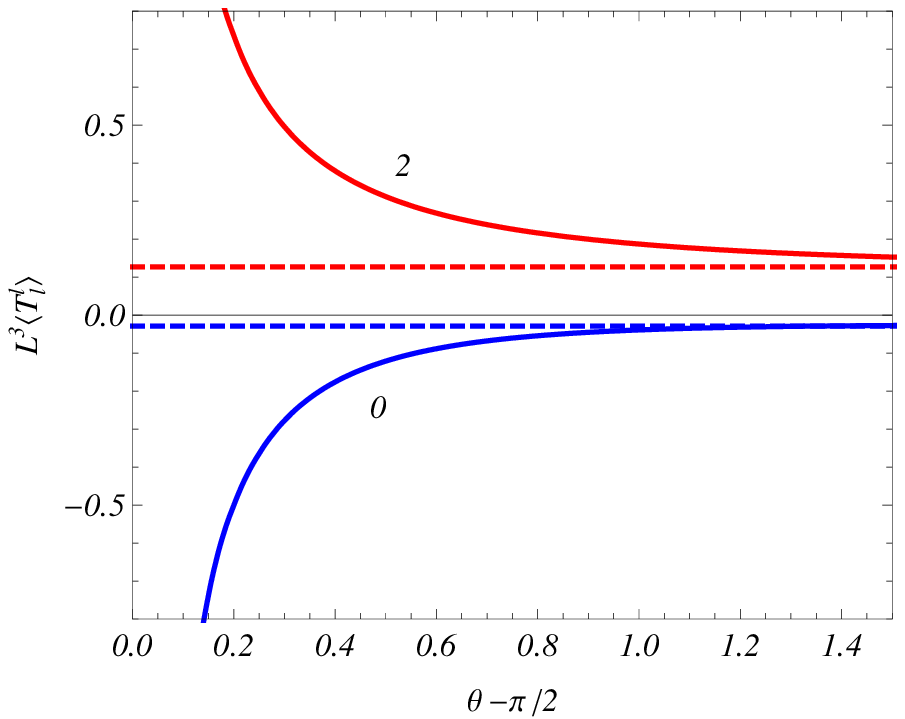,width=7.cm,height=5.5cm} & \quad %
\epsfig{figure=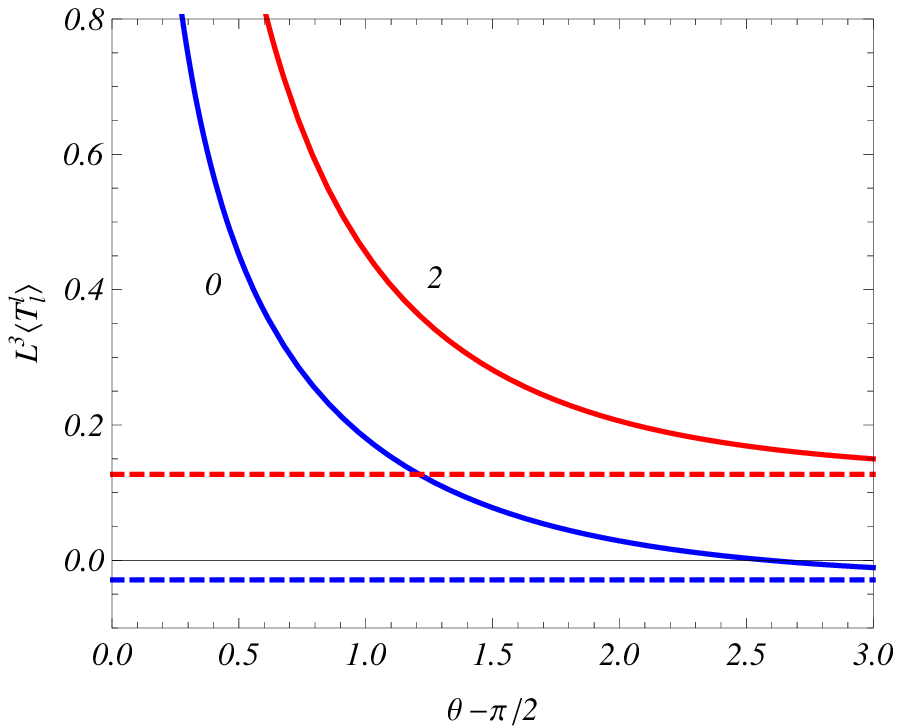,width=7.cm,height=5.5cm}%
\end{tabular}%
\end{center}
\caption{The components of the vacuum energy-momentum tensor on the tube,
versus the distance from the boundary, for conformally (left panel) and
minimally (right panel) coupled scalar fields. The numbers near the curves
are the values of the index $l$. The dashed lines are the corresponding
components for an infinite tube. The graphs are plotted for $ma=1/2$. }
\label{fig3}
\end{figure}

In figure \ref{fig3} we have displayed the components of the vacuum
energy-momentum tensor on the tube as functions of the distance from the
boundary $\theta =\pi /2$. The numbers near the curves correspond to the
value of the index $l$. The left/right panels correspond to
conformally/minimally coupled scalar fields. The dashed curves present the
same quantities in the geometry of an infinite tube (recall that they do not
depend on the curvature coupling parameter). The graphs are plotted for $%
ma=1/2$. For a conformally coupled field the cap-induced part is positive
for the energy density and negative for the $_{2}^{2}$-stress. For a
minimally coupled field the cap-induced contributions are positive for both
these components. The cup-induced contribution in the $_{1}^{1}$-stress
vanishes and the corresponding graphs coincide with the dashed lines in
figure \ref{fig3} for the energy density.

\subsection{Energy-momentum tensor on the cap}

Now let us consider the energy-momentum tensor on the hemisphere cap, $%
0\leqslant \theta <\pi /2$. By using the formula (\ref{EMT}), after long
calculations, the corresponding nonzero components are expressed in the form%
\begin{equation}
\langle T_{l}^{l}\rangle =\langle T_{l}^{l}\rangle _{S^{2}}+\frac{a^{-3}}{%
4\pi }\sideset{}{'}{\sum}_{n=0}^{\infty }\int_{ma}^{\infty }dy\,\frac{%
yg_{n}(y)}{\sqrt{y^{2}-m^{2}a^{2}}}\frac{G_{n}^{(l)}(z(y),\cos \theta )}{%
\cosh [\pi z(y)]},  \label{Tllcap}
\end{equation}%
where $\langle T_{l}^{l}\rangle _{S^{2}}$ is the renormalized VEV on a
2-dimensional sphere. The functions in the integrand of (\ref{Tllcap}) for
separate components are defined by the expressions (with no summation over $%
l $)%
\begin{eqnarray}
G_{n}^{(0)}(x,u) &=&\left( 4\xi -1\right) \left\{
(1-u^{2})[P_{ix-1/2}^{-n\prime }(u)]^{2}\right.  \notag \\
&&\left. +\left( x^{2}-\frac{3}{4}+\frac{n^{2}}{1-u^{2}}+\frac{%
x^{2}-m^{2}a^{2}-1/4}{2\xi -1/2}\right) [P_{ix-1/2}^{-n}(u)]^{2}\right\} ,
\notag \\
G_{n}^{(1)}(x,u) &=&(1-u^{2})[P_{ix-1/2}^{-n\prime }(u)]^{2}-2\xi u\partial
_{u}[P_{ix-1/2}^{-n}(u)]^{2}  \notag \\
&&-\left( \frac{1}{4}-2\xi +x^{2}+\frac{n^{2}}{1-u^{2}}\right)
[P_{ix-1/2}^{-n}(u)]^{2},  \notag \\
G_{n}^{(2)}(x,u) &=&\left( 4\xi -1\right) (1-u^{2})[P_{ix-1/2}^{-n\prime
}(u)]^{2}+2\xi u\partial _{u}[P_{ix-1/2}^{-n}(u)]^{2}  \notag \\
&&+\left[ \left( 4\xi -1\right) \left( \frac{1}{4}+x^{2}\right) +n^{2}\frac{%
4\xi +1}{1-u^{2}}+2\xi \right] [P_{ix-1/2}^{-n}(u)]^{2},  \label{Gn2}
\end{eqnarray}%
where $P_{ix-1/2}^{-n\prime }(u)=\partial _{u}P_{ix-1/2}^{-n}(u)$ and we
have used the differential equation for the associated Legendre function to
exclude the second derivative of this function. By using the equation for
the function $P_{ix-1/2}^{n}(u)$, the following relation can be proved:%
\begin{equation}
\left( 1-u^{2}\right) \partial
_{u}G_{n}^{(1)}(x,u)=u[G_{n}^{(1)}(x,u)-G_{n}^{(2)}(x,u)].  \label{RelG1}
\end{equation}

The second term in the rhs of (\ref{Tllcap}) is induced by the cylindrical
geometry. For points outside the boundary, $\theta <\pi /2$, the
renormalization is required for the part $\langle T_{l}^{l}\rangle _{S^{2}}$
only. The corresponding procedure is described in appendix \ref{sec:App2}
and the renormalized energy density and stresses are given by the
expressions (\ref{T00S22}) and (\ref{T11S21}). The special case of a
conformally coupled field has been discussed in \cite{Bord09}.

At $\theta =0$ the only nonzero contribution to the tube-induced part in (%
\ref{Tllcap}) comes from the terms $n=0$ and $n=1$. By using the expressions
for the function $P_{ix-1/2}^{-n}(u)$ at $u=1$ it can be seen that the
stresses are isotropic at $\theta =0$: $\langle T_{1}^{1}\rangle =\langle
T_{2}^{2}\rangle $.

We can check that the tube-induced part in (\ref{Tllcap}) obeys the trace
relation (\ref{TraceRel}). In particular, for a conformally coupled massless
field the vacuum energy-momentum tensor id traceless. As is well-known, in
odd spacetime dimensions the trace anomaly is absent. As an additional
check, we can see that the tube-induced contribution to the VEV of the
energy-momentum tensor obeys the covariant conservation equation $\nabla
_{k}\langle T_{i}^{k}\rangle =0$. On the cap it is reduced to a single
equation%
\begin{equation}
\partial _{\theta }\langle T_{1}^{1}\rangle +\cot \theta \left( \langle
T_{1}^{1}\rangle -\langle T_{2}^{2}\rangle \right) =0.  \label{ConsEq}
\end{equation}%
The validity of this equation for the tube-induced part is directly obtained
by using the relation (\ref{RelG1}).

Now let us consider the asymptotic behavior of the tube-induced contribution
in (\ref{Tllcap}) near the boundary $\theta =\pi /2$. The energy density and
the stress $\langle T_{2}^{2}\rangle $ diverge on the boundary. For points
close to the boundary the dominant contribution in the tube-induced part is
given by large values of $n$ and $y$. Introducing a new integration variable
$x=y/n$, we use the uniform asymptotic expansions for the associated
Legendre function and its derivative \cite{Khus03}. For $|u|\ll 1$ the
leading term is given by the expression%
\begin{equation}
P_{iz(nx)-1/2}^{-n}(u)\sim \frac{e^{n-n\ln n}}{\sqrt{2\pi n}}\frac{%
e^{n(x\arctan x-u\sqrt{1+x^{2}})}}{\left( 1+x^{2}\right) ^{n/2+1/4}},
\label{Pas}
\end{equation}%
and for the derivative one has $P_{iz(nx)-1/2}^{-n\prime }(u)\sim -n\sqrt{%
1+x^{2}}P_{iz(nx)-1/2}^{-n}(u)$. By using the asymptotic formula for the
gamma function and the relation (\ref{fnas}), we can also see that%
\begin{eqnarray}
g_{n}(nx) &\approx &\left( 1+x^{2}\right) ^{n-2}\frac{e^{-2n+2n\ln n+\pi nx}%
}{8n^{2}e^{2nx\arctan x}}  \notag \\
&&\times \left[ (1+x^{2})\left( 1-4\xi \right) -1\right] .  \label{gnas}
\end{eqnarray}%
Substituting (\ref{Pas}) and (\ref{gnas}) into the tube-induced part in (\ref%
{Tllcap}) we get (no summation over $l$)
\begin{equation}
\langle T_{l}^{l}\rangle \approx -\frac{2\xi \left( 4\xi -1\right) +(l+1)/16%
}{16\pi (\pi -2\theta )a^{3}},  \label{T00near}
\end{equation}%
where $l=0,2$. As is seen, the leading terms in the asymptotic expansions
for the energy density and the stress $\langle T_{2}^{2}\rangle $ on the cup
and on the tube, as functions of $2\theta -\pi $, coincide. For the stress $%
\langle T_{1}^{1}\rangle $ the leading term in the asymptotic expansion
vanishes. For this component it is more convenient to use the equation (\ref%
{ConsEq}) and (\ref{T00near}). From the latter we see that the stress $%
\langle T_{1}^{1}\rangle $ is finite on the boundary. By using the
asymptotic expressions (\ref{T00near}) and the conservation equation (\ref%
{ConsEq}), for the normal derivative of the radial stress at the boundary
one has%
\begin{equation}
\partial _{\theta }T_{1}^{1}|_{\theta =\pi /2-0}=-\frac{2\xi \left( 4\xi
-1\right) +3/16}{32\pi a^{3}}.  \label{DerT11}
\end{equation}%
As will be seen below, the $_{1}^{1}$-stress is continuous at the boundary.
By taking into account that $\partial _{\theta }T_{1}^{1}|_{\theta =\pi
/2+0}=0$, we see that its normal derivative has discontinuity.

\begin{figure}[tbph]
\begin{center}
\begin{tabular}{cc}
\epsfig{figure=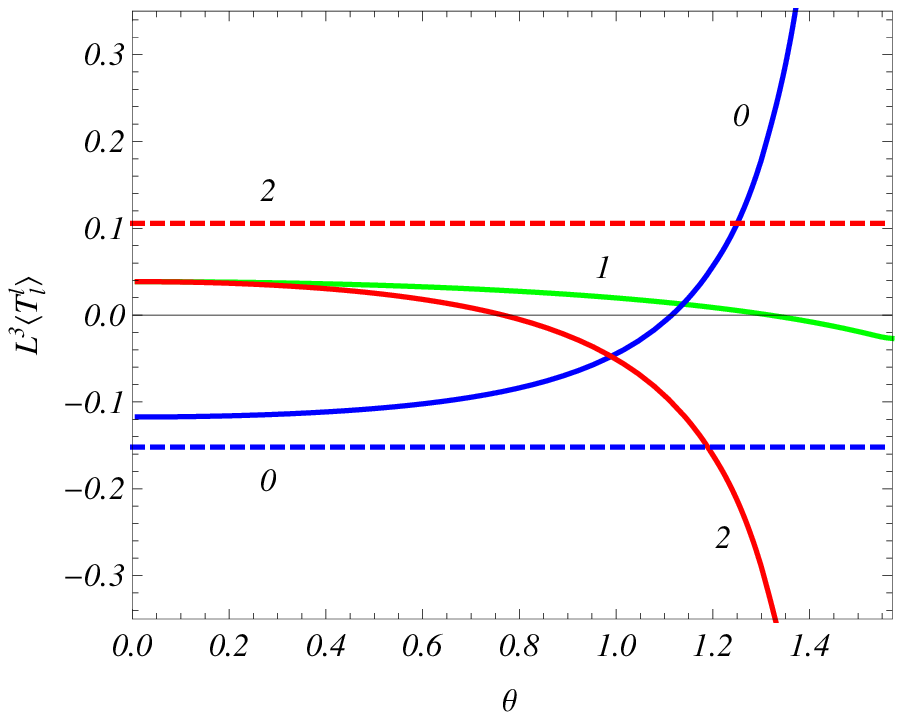,width=7.cm,height=5.5cm} & \quad %
\epsfig{figure=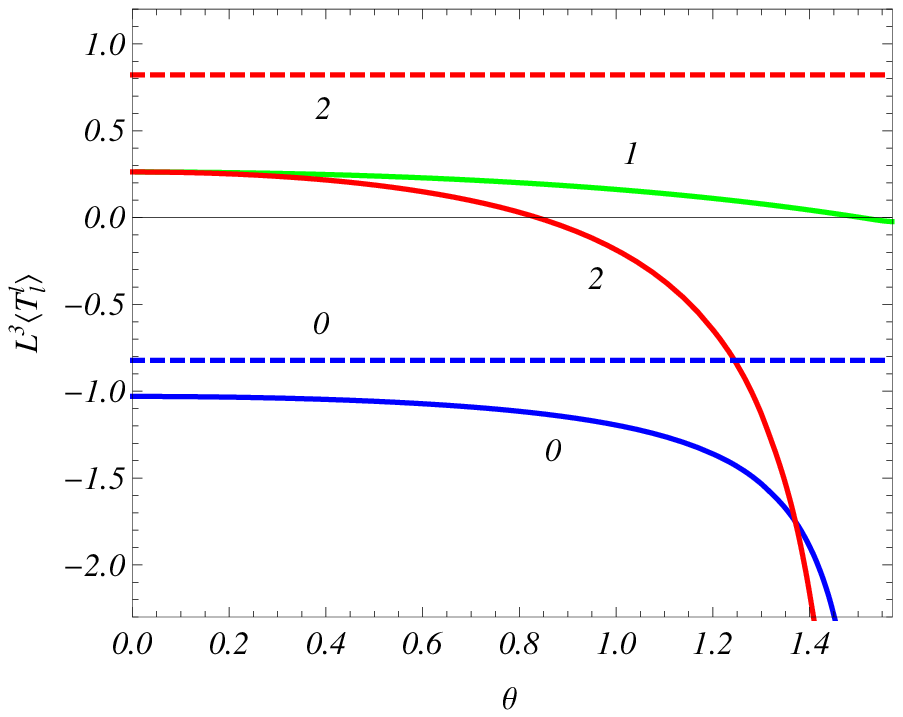,width=7.cm,height=5.5cm}%
\end{tabular}%
\end{center}
\caption{The vacuum energy density and stresses on the hemispherical cap as
functions of $\protect\theta $ for conformally (right panel) and minimally
(right panel) coupled fields. The numbers near the curves are the values of
the index $l$. The dashed lines correspond to the vacuum densities on $S^{2}$%
. The graphs are plotted for $ma=1/2$. }
\label{fig4}
\end{figure}

Figure \ref{fig4} presents the components of the vacuum energy-momentum
tensor (the numbers near the graphs correspond to the values of the index $l$%
) on the hemisphere cap as functions of the angle $\theta $. The dashed
curves present the same quantities for the background geometry $S^{2}$. The
left and right panels correspond to conformally and minimally coupled scalar
fields, respectively. As before, the graphs are plotted for $ma=1/2$. For a
conformally coupled field, the tube-induced part of the energy density is
positive and the $_{2}^{2}$-stress is negative. For a minimally coupled
field, the tube induced contributions are negative.

\subsection{The Casimir force}

The Casimir force acting per unit length of the boundary $\theta =\pi /2$
(the Casimir pressure) is determined by the normal stress evaluated at the
boundary:%
\begin{equation}
p=-\langle T_{1}^{1}\rangle _{\theta =\pi /2}.  \label{Force}
\end{equation}%
For the side of the boundary corresponding to the tube, $\theta =\pi /2+0$,
the cap-induced part in the normal stress vanishes and the Casimir pressure
is directly obtained from (\ref{Tll0}):
\begin{equation}
p_{+}=\frac{\mathrm{Li}_{3}(e^{-mL})+mL\,\mathrm{Li}_{2}(e^{-mL})}{2\pi L^{3}%
}.  \label{p+}
\end{equation}%
This pressure is positive and the corresponding force is directed to the
direction of the cap. For a massless field%
\begin{equation}
p_{+}=p_{0}=\frac{\zeta (3)}{2\pi L^{3}}\approx \frac{0.1913}{L^{3}}.
\label{p0}
\end{equation}

The pressure from the cap side is determined from (\ref{Tllcap}) with $l=1$
evaluated at $\theta =\pi /2-0$. It is expresses in the decomposed form%
\begin{equation}
p_{-}=p_{S^{2}}+p_{\mathrm{ind}},  \label{p-1}
\end{equation}%
where $p_{S^{2}}=-\langle T_{1}^{1}\rangle _{S^{2}}$ and the part
\begin{eqnarray}
p_{\mathrm{ind}} &=&\frac{a^{-3}}{4\pi ^{2}}\sideset{}{'}{\sum}%
_{n=0}^{\infty }\int_{ma}^{\infty }dy\,\frac{y}{\sqrt{y^{2}-m^{2}a^{2}}}
\notag \\
&&\times \frac{\lbrack \sqrt{y^{2}+n^{2}}/P_{iz(y)-1/2}^{-n\prime
}+1/P_{iz(y)-1/2}^{-n}]^{2}}{\Gamma (n+iz(y)+1/2)\Gamma (n-iz(y)+1/2)},
\label{pind}
\end{eqnarray}%
is induced by the geometry of the tube. In deriving this formula we have
used the expressions (\ref{P0}) and the relation (\ref{RelP0}). The
numerical calculations show that the Casimir pressures from different sides
of the boundary are equal to each other, i.e., $p_{+}=p_{-}$, and, hence,
the net Casimir force on the boundary vanishes. This is related to the
smooth transition between two subspaces and to the fact that the extrinsic
curvature tensor of the separating boundary vanishes for both sides.

\begin{figure}[tbph]
\begin{center}
\epsfig{figure=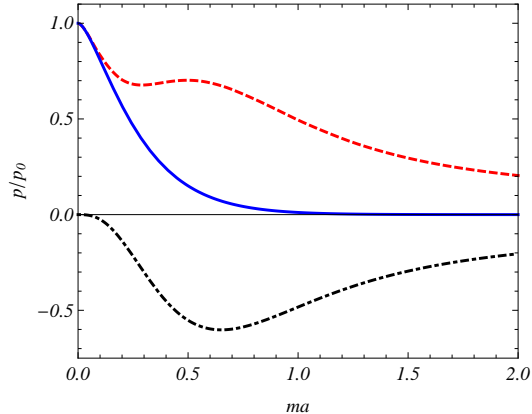,width=7.cm,height=5.5cm}
\end{center}
\caption{The ratio of the Casimir pressure on the separating boundary to the
corresponding pressure for a massless field, as a function of $ma$. The
dot-dashed curve corresponds to the ratio $p_{S^{2}}/p_{0}$ and the dashed
curve is for $p_{\mathrm{ind}}/p_{0}$. }
\label{fig5}
\end{figure}

In figure \ref{fig5}, for a conformally coupled scalar field, we have
plotted the ratio of the Casimir pressure $p=p_{-}=p_{+}$ to $p_{0}$,
defined by (\ref{p0}), as a function of $ma$. We have also presented the
parts corresponding to the separate terms in the rhs of (\ref{p-1}). The
dot-dashed curve corresponds to the ratio $p_{S^{2}}/p_{0}$ (the part
corresponding to the pressure in the geometry of $S^{2}$) and the dashed
curve corresponds to the ratio for the tube-induced part, $p_{\mathrm{ind}%
}/p_{0}$. Note that for large values of the mass, the decay of the separate
terms in the rhs of (\ref{p-1}), as functions of $ma$, is as power-law,
whereas the total pressure decays exponentially. The latter follows from the
equality $p_{+}=p_{-}$ and from the exponential decay of $p_{+}$ for large
masses.

\section{Conclusion}

\label{sec:Conc}

In the present paper we have considered the properties of the quantum vacuum
for a $(2+1)$-dimensional scalar field in a background geometry of a
cylindrical tube with a hemispherical cap. In this geometry one has two
spatial regions with different geometrical characteristics separated by a
circular boundary. Our main interest was to investigate the changes in the
local characteristics of the vacuum induced by the geometry of the attached
region. In free field theories with only interaction of the background
gravitational field, all the information about the properties of the vacuum
state is encoded in two-point functions. As such we have evaluated the
positive frequency Wightman function by using the direct summation method
over a complete set of modes. In the problem under consideration the metric
tensor and its first derivatives are continuous at the separating boundary
and, as a consequence, the mode functions and their normal derivatives are
continuous as well. For continuous energy spectrum these functions in
separate regions are given by (\ref{phi2S2}) and (\ref{phi2}) with the
relations (\ref{ealf}) and (\ref{Cs}) between the constants. In addition,
for negative values of the curvature coupling parameter $\xi $ there are
bound states. In both the cylindrical and hemispherical subspaces, we
explicitly separated the contributions in the Wightman function induced by
the geometry of the attached region. These contributions are given by the
second term in the rhs of (\ref{WFcyl2}) for points on the tube and by the
second term in the rhs of (\ref{WFcap3}) for the hemispherical cap. With
this separation, the renormalization of the VEVs in the coincidence limit is
reduced to that for an infinite cylinder and for two-dimensional sphere $%
S^{2}$.

As an important local characteristic of the vacuum state, in section \ref%
{sec:phi2} we have studied the VEV of the field squared. On the tube this
VEV is given by (\ref{phi2cyl}), where the second term in the rhs is the
contribution induced by the hemispherical cap. Unlike to the case of a
semi-infinite tube with Dirichlet or Neumann boundary conditions on its
edge, where the mean field squared diverges on the boundary, in the problem
at hand the VEV of the field squared is finite everywhere. At large
distances from the separating boundary and for a massive field, the
contribution in the mean field squared induced by the cap is exponentially
small (see (\ref{phi2cyllarge})) and the total VEV is dominated by the part
corresponding to an infinite tube. The latter is positive everywhere. For
points on the hemispherical cap the VEV of the field squared is presented in
the decomposed form (\ref{phi2cap}). The tube-induced part, given by the
second term in the rhs, is finite everywhere including the points on the
boundary. Though the Wightman function is continuous at the separating
circle, the renormalized VEV of the field squared has discontinuity. This is
related to the fact that in the renormalization procedure for tubular and
hemispherical geometries different terms are subtracted. On the hemisphere,
in addition to the Minkowskian term, we have also subtracted a finite
renormalization term (see appendix \ref{sec:App1}). In the minimal
subtraction scheme, where the only Minkowskian part is subtracted for both
regions, the mean field squared is continuous at the boundary. In the
numerical examples, we have considered the most important special cases of
minimally and conformally coupled scalar fields. For both these cases the
cap-induced contributions in the VEV of the field squared on the tube are
positive, whereas the tube-induced contributions on the cap are negative.

Another important characteristic of the ground state is the VEV of the
energy-momentum tensor. This VEV is diagonal and obeys the trace relation (%
\ref{TraceRel}). On the tube the vacuum energy-momentum tensor is given by
the expression (\ref{Tllcyl}) with the infinite tube part from (\ref{Tll0}).
The cap-induced part in the vacuum stress normal to the boundary vanishes.
The energy density and the parallel stress diverge on the boundary. The
leading terms in the corresponding asymptotic expansion over the distance
from the boundary are given by (\ref{NearB}). In the geometry of a
semi-infinite tube with Dirichlet or Neumann boundary conditions on the
edge, the divergences are stronger, the VEVs diverge as the inverse cube of
the distance from the edge. At large distances from the boundary and for a
massive field, the cup-induced contribution in the energy-momentum tensor is
exponentially small, whereas for a massless field it decays as the inverse
square of the distance. For numerical examples displayed in figure \ref{fig3}%
, the cup-induced contribution to the energy density is positive for both
conformally and minimally coupled field. The corresponding parallel stress
is negative for a conformally coupled field and positive for a minimally
coupled field.

The vacuum energy-momentum tensor on the cap is decomposed as (\ref{Tllcap}%
), where the components for the part corresponding to $S^{2}$ are given by (%
\ref{T00S22}) and (\ref{T11S21}). We have explicitly checked that the VEV
obeys the covariant conservation equation, which for the geometry of the cup
is reduced to the relation (\ref{ConsEq}). The energy density and the
parallel stress diverge on the separating circle and the expression for the
corresponding leading terms in the asymptotic expansion over $\theta -\pi /2$
coincide with those on the tube. On the cap and on the tube these components
have opposite signs. The normal stress is finite everywhere. It determines
the Casimir force acting on the boundary. From the sides of the tube and of
the cap the vacuum pressures on the boundary are given by the expressions (%
\ref{p+}) and (\ref{p-1}) respectively. We have checked numerically that
these pressures coincide and, hence, the net Casimir force on the boundary
is zero.

The results obtained above may be applied to carbon nanotubes with
half fullerene caps, described in the long-wavelength approximation
by an effective field theory. As has been mentioned in Introduction,
the latter, in addition to Dirac fermions, involves scalar and gauge
fields as well.

\section*{Acknowledgments}

The authors thank Conselho Nacional de Desenvolvimento Cient\'{\i}fico e
Tecnol\'{o}gico (CNPq) for the financial support. A. A. S. was supported by
the State Committee of Science of the Ministry of Education and Science RA,
within the frame of Grant No. SCS 13-1C040.

\appendix

\section{Proof of the identity with Legendre functions}

\label{sec:App1}

In this appendix we prove the identity which is used in section \ref{sec:WF}
for the decomposition of the Wightman function on the hemispherical cap. Let
us define the function%
\begin{equation}
R_{\lambda -1/2}^{\mu }(u)=Q_{-\lambda -1/2}^{\mu }(u)\cos [\pi (\mu
-\lambda )]+Q_{\lambda -1/2}^{\mu }(u)\cos [\pi (\mu +\lambda )],  \label{R}
\end{equation}%
where $u=\cos \theta $ and $Q_{\lambda -1/2}^{\mu }(u)$ is the associated
Legendre function of the second kind. By using the relation \cite{Erde53}
\begin{equation}
Q_{-\lambda -1/2}^{\mu }(u)\cos [\pi (\mu -\lambda )]=Q_{\lambda -1/2}^{\mu
}(u)\cos [\pi (\mu +\lambda )]+\pi \cos (\mu \pi )\sin (\lambda \pi
)P_{\lambda -1/2}^{\mu }(u),  \label{Rel3}
\end{equation}%
the function (\ref{R}) can also be expressed as a linear combination of the
functions $P_{\lambda -1/2}^{\mu }(u)$ and $Q_{\lambda -1/2}^{\mu }(u)$.
Next, we introduce the notations%
\begin{eqnarray}
\bar{P}_{\lambda -1/2}^{\pm ,\mu }(u) &=&\partial _{\theta }P_{\lambda
-1/2}^{\mu }(u)\mp ipP_{\lambda -1/2}^{\mu }(u),  \notag \\
\bar{R}_{\lambda -1/2}^{\pm ,\mu }(u) &=&\partial _{\theta }R_{\lambda
-1/2}^{\mu }(u)\mp ipR_{\lambda -1/2}^{\mu }(u).  \label{BarNot}
\end{eqnarray}%
We want to prove the identity%
\begin{equation}
\frac{\cos [\pi (\mu +\lambda )]}{\bar{P}_{\lambda -1/2}^{+,\mu }(u)\bar{P}%
_{\lambda -1/2}^{-,\mu }(u)}=i\frac{\sin \theta }{4p}\frac{\Gamma (\lambda
-\mu +1/2)}{\Gamma (\lambda +\mu +1/2)}\sum_{j=+,-}j\frac{\bar{R}_{\lambda
-1/2}^{j,\mu }(u)}{\bar{P}_{\lambda -1/2}^{j,\mu }(u)}.  \label{Ident1}
\end{equation}

For the sum in the rhs of (\ref{Ident1}) one has%
\begin{equation}
\sum_{j=+,-}j\frac{\bar{R}_{\lambda -1/2}^{j,\mu }(u)}{\bar{P}_{\lambda
-1/2}^{j,\mu }(u)}=-2ip\frac{R_{\lambda -1/2}^{\mu }(u)\partial _{\theta
}P_{\lambda -1/2}^{\mu }(u)-P_{\lambda -1/2}^{\mu }(u)\partial _{\theta
}R_{\lambda -1/2}^{\mu }(u)}{\bar{P}_{\lambda -1/2}^{+,\mu }(u)\bar{P}%
_{\lambda -1/2}^{-,\mu }(u)}.  \label{Rel1}
\end{equation}%
By using the Wronskian for the associated Legendre functions we can see that
\begin{equation}
Q_{\lambda -1/2}^{\mu }(u)\partial _{\theta }P_{\lambda -1/2}^{\mu
}(u)-P_{\lambda -1/2}^{\mu }(u)\partial _{\theta }Q_{\lambda -1/2}^{\mu }(u)=%
\frac{\Gamma (\lambda +\mu +1/2)}{\sin \theta \Gamma (\lambda -\mu +1/2)}.
\label{Rel2}
\end{equation}%
Combining this with the relation (\ref{Rel3}), for the numerator in (\ref%
{Rel1}) we get%
\begin{equation}
R_{\lambda -1/2}^{\mu }(u)\partial _{\theta }P_{\lambda -1/2}^{\mu
}(u)-P_{\lambda -1/2}^{\mu }(u)\partial _{\theta }R_{\lambda -1/2}^{\mu }(u)=%
\frac{2\Gamma (\lambda +\mu +1/2)}{\sin \theta \Gamma (\lambda -\mu +1/2)}%
\cos [\pi (\mu +\lambda )].  \label{Rel4}
\end{equation}%
The identity (\ref{Ident1}) directly follows from (\ref{Rel1}) and (\ref%
{Rel4}).

For $\mu =-n$ from (\ref{Ident1}) one has%
\begin{equation}
\frac{1}{\bar{P}_{\lambda -1/2}^{+,-n}(u)\bar{P}_{\lambda -1/2}^{-,-n}(u)}=i%
\frac{\sin \theta }{4p}\frac{\Gamma (\lambda +n+1/2)}{\Gamma (\lambda -n+1/2)%
}\sum_{j=+,-}j\frac{\bar{S}_{\lambda -1/2}^{j,-n}(u)}{\bar{P}_{\lambda
-1/2}^{j,-n}(u)},  \label{Ident2}
\end{equation}%
where we have defined a new function%
\begin{equation}
S_{\lambda -1/2}^{-n}(u)=Q_{\lambda -1/2}^{-n}(u)+Q_{-\lambda -1/2}^{-n}(u),
\label{S}
\end{equation}%
and, similar to (\ref{BarNot}),
\begin{equation}
\bar{S}_{\lambda -1/2}^{\pm ,\mu }(u)=\partial _{\theta }S_{\lambda
-1/2}^{\mu }(u)\mp ipS_{\lambda -1/2}^{\mu }(u).  \label{Sbar}
\end{equation}%
Note that, by using the relation (\ref{Rel3}), the function (\ref{S}) can
also be written in the form%
\begin{equation}
S_{\lambda -1/2}^{-n}(x)=2Q_{\lambda -1/2}^{-n}(u)+\pi \tan (\lambda \pi
)P_{\lambda -1/2}^{-n}(u).  \label{S2}
\end{equation}

By making use of the trigonometric expansions of the functions $P_{\lambda
-1/2}^{-n}(u)$ and $Q_{\lambda -1/2}^{-n}(u)$ \cite{Erde53}, the following
expansion is obtained for the function (\ref{S}):%
\begin{eqnarray}
S_{\lambda -1/2}^{-n}(u) &=&\frac{2^{1-n}}{\sin ^{n}\theta }\frac{\sqrt{\pi }%
\Gamma (\lambda -n+1/2)}{\Gamma (\lambda +1)\cos (\lambda \pi )}%
\sum_{l=0}^{\infty }\frac{(1/2-n)_{l}(\lambda -n+1/2)_{l}}{l!(\lambda +1)_{l}%
}  \notag \\
&&\times \cos \left[ \left( 2l-n+1/2\right) \theta -\lambda \left( \pi
-\theta \right) \right] ,  \label{Sexp}
\end{eqnarray}%
where $(b)_{l}$ is Pochhammer's symbol, $u=\cos \theta $ and $0<\theta <\pi $%
. From here it follows that in the complex plane $\lambda $ the function $%
S_{\lambda -1/2}^{-n}(u)$ decays exponentially in the limit $|\mathrm{Im}%
\,\lambda |\rightarrow \infty $. Note that in the same limit the function $%
P_{\lambda -1/2}^{-n}(u)$ increases exponentially.

\section{Casimir densities on $S^{2}$}

\label{sec:App2}

In this appendix we consider the vacuum densities for a scalar field on $%
S^{2}$ (for the Casimir energy in spherical universes see \cite{Ford75} and
references therein). The corresponding vacuum energy and stresses for a
conformally coupled field have been investigated in \cite{Bord09}. The mode
functions on $S^{2}$, regular for $0\leqslant \theta <\pi $, have the form (%
\ref{phi2S2}). From the regularity of these functions at $\theta =\pi $ it
follows that $\lambda -1/2=l$ with $l=|n|,|n|+1,\ldots $. Hence, the modes
are specified by the set $\sigma =(l,n)$ and have the form%
\begin{equation}
\varphi _{\sigma }(x)=C_{0\mathrm{s}}P_{l}^{-|n|}(\cos \theta )e^{in\phi
-i\omega _{l}t},\;0\leqslant \theta \leqslant \pi ,  \label{phiS2}
\end{equation}%
where $|n|\leqslant l$, $l=0,1,2,\ldots $, and the energy $\omega _{l}$ is
given by the expression (\ref{oml}). From the normalization condition one
gets%
\begin{equation}
|C_{0\mathrm{s}}|^{2}=\frac{l+1/2}{4\pi a^{2}\omega _{l}}\frac{\Gamma
(l+|n|+1)}{\Gamma (l-|n|+1)}.  \label{C0s}
\end{equation}%
Substituting the functions (\ref{phiS2}) into the mode-sum, after the
summation over $n$ by using (\ref{SumForm}), we obtain the expression (\ref%
{WS2}) for the Wightman function on $S^{2}$. Note that the latter obeys the
relation%
\begin{equation}
\partial _{a}[a^{3}\lim_{t^{\prime }\rightarrow t}\partial _{t^{\prime
}}\partial _{t}W_{S^{2}}(x,x^{\prime })]=m^{2}a^{2}\lim_{t^{\prime
}\rightarrow t}W_{S^{2}}(x,x^{\prime }).  \label{RelWFS2}
\end{equation}%
As will be seen below, this leads to the relation between the vacuum energy
and the VEV of the field squared.

For the VEV of the field squared we find%
\begin{equation}
\langle \varphi ^{2}\rangle _{S^{2}}=\frac{a^{-2}}{4\pi }\sum_{l=0}^{\infty }%
\frac{l+1/2}{\omega _{l}}.  \label{phi2S21}
\end{equation}%
Of course this expression is divergent and needs to be renormalized. Here we
note that in $(2+1)$-dimensions the divergence of the Wightman function in
the coincidence limit is the same as in Minkowski spacetime. Hence, the
subtraction of the Minkowskian part from (\ref{phi2S21}) gives a finite
result. The unrenormalized VEV of the field squared in the Minkowski
spacetime is given by the expression%
\begin{equation}
\langle \varphi ^{2}\rangle _{M}=\frac{1}{4\pi }\int_{0}^{\infty }dx\,\frac{x%
}{\sqrt{x^{2}+m^{2}}}.  \label{phi2M}
\end{equation}%
However, in the renormalization procedure, we need also to subtract the next
to the leading term in DeWitt-Schwinger expansion.

First we assume that $m^{2}a^{2}+2\xi \geqslant 1/4$. In this case, applying
the Abel-Plana summation formula in the form \cite{Bord09,Saha07Rev}
\begin{equation}
\sum_{l=0}^{\infty }f(l+1/2)=\int_{0}^{\infty }dx\,f(x)-i\int_{0}^{\infty
}dx\,\frac{f(ix)-f(-ix)}{e^{2\pi x}+1},  \label{AP}
\end{equation}%
and subtracting the first two terms of DeWitt-Schwinger expansion, from (\ref%
{phi2S21}) for the renormalized VEV of the field squared we get%
\begin{equation}
\langle \varphi ^{2}\rangle _{S^{2}}=\frac{m}{4\pi }\left[ 1+\frac{\xi -1/6}{%
m^{2}a^{2}}-b_{m}+2b_{m}\int_{0}^{1}dx\,\frac{x(1-x^{2})^{-1/2}}{e^{2\pi
\omega _{m}x}+1}\right] ,  \label{phi2ren}
\end{equation}%
with the notation $b_{m}=\omega _{m}/(ma)$. The second term in the square
brackets of (\ref{phi2ren}) comes from the next to the leading term in
DeWitt-Schwinger expansion. For large masses, $ma\gg 1$, one has%
\begin{equation}
\langle \varphi ^{2}\rangle _{S^{2}}\approx \frac{m^{-3}}{2^{9}\pi a^{4}}%
\left[ \frac{7}{15}+\left( 8\xi -1\right) ^{2}\right] ,
\label{phi2renlargem}
\end{equation}%
and the VEV decays as $m^{-3}$.

The VEV of the energy-momentum tensor is evaluated by using the formula (\ref%
{EMT}) with the Wightman function from (\ref{WS2}). For the unrenormalized
energy density we get%
\begin{equation}
\langle T_{0}^{0}\rangle _{S^{2}}=\frac{a^{-3}}{4\pi }\sum_{l=0}^{\infty
}(l+1/2)\sqrt{(l+1/2)^{2}+\omega _{m}^{2}},  \label{T00S2}
\end{equation}%
and for the stresses one has the relation%
\begin{equation}
\langle T_{1}^{1}\rangle _{S^{2}}=\langle T_{2}^{2}\rangle _{S^{2}}=-\frac{1%
}{2}\langle T_{0}^{0}\rangle _{S^{2}}+\frac{1}{2}m^{2}\langle \varphi
^{2}\rangle _{S^{2}}.  \label{T11S2}
\end{equation}
Note that the VEVs obey the trace relation (\ref{TraceRel}). For the
renormalization we need to subtract from the VEVs the corresponding
DeWitt-Schwinger expansion truncated at the adiabatic order 3.

An alternative way is to use the relation $\partial _{a}\left( a^{3}\langle
T_{0}^{0}\rangle _{S^{2}}\right) =m^{2}a^{2}\langle \varphi ^{2}\rangle
_{S^{2}}$, which directly follows from (\ref{RelWFS2}). From this relation
for the renormalized VEV of the energy density we get
\begin{equation}
\langle T_{0}^{0}\rangle _{S^{2}}=\frac{m^{2}}{a^{3}}\int da\,a^{2}\langle
\varphi ^{2}\rangle _{S^{2}}.  \label{T00S21}
\end{equation}%
By taking into account the expression (\ref{phi2ren}), one finds%
\begin{equation}
\langle T_{0}^{0}\rangle _{S^{2}}=\frac{m^{3}}{4\pi }\,\left( \frac{1}{3}+%
\frac{\xi -1/6}{m^{2}a^{2}}-\frac{b_{m}^{3}}{3}+2b_{m}^{3}\int_{0}^{1}dx\,%
\frac{x\sqrt{1-x^{2}}}{e^{2\pi \omega _{m}x}+1}\right) .  \label{T00S22}
\end{equation}%
For large values of the mass we have the leading term%
\begin{equation}
\langle T_{0}^{0}\rangle _{S^{2}}\approx -\frac{a^{-4}}{8\pi m}\,\left[
\frac{7}{960}+\left( \xi -\frac{1}{8}\right) \left( \xi -\frac{5}{24}\right) %
\right] ,  \label{T00S2m}
\end{equation}%
and the renormalized VEV vanishes in the limit $m\rightarrow \infty $. Note
that the relation (\ref{T00S21}) determines the renormalized energy density
up to the term $C(m)/a^{3}$, where the function $C(m)$ is dimensionless. By
taking into account that the only dimensionful quantity to form this
function is the mass $m$, we conclude that $C(m)$ does not depend on $m$.
Now, imposing the renormalization condition (for a discussion of this
condition see \cite{Bord09}) $\langle T_{0}^{0}\rangle _{S^{2}}\rightarrow 0$
in the limit $m\rightarrow \infty $, we see that $C(m)=0$. The stresses are
found from the relation (\ref{T11S2}):%
\begin{eqnarray}
\langle T_{1}^{1}\rangle _{S^{2}} &=&\frac{m^{3}}{8\pi }\left\{
2b_{m}^{3}\int_{0}^{1}dx\,\frac{x^{3}/\sqrt{1-x^{2}}}{e^{2\pi \omega _{m}x}+1%
}-\left( b_{m}-\,1\right) \right.  \notag \\
&&\times \left. \left[ \frac{2}{3}+b_{m}\left( b_{m}+1\right) \left(
2\int_{0}^{1}dx\,\frac{x/\sqrt{1-x^{2}}}{e^{2\pi \omega _{m}x}+1}-\frac{1}{3}%
\right) \right] \right\} .  \label{T11S21}
\end{eqnarray}%
In the limit of large masses, to the leading order one has $\langle
T_{1}^{1}\rangle _{S^{2}}\approx -\langle T_{0}^{0}\rangle _{S^{2}}$ with
the asymptotic expression for the energy density given by (\ref{T00S2m}).
For a conformally coupled scalar field $b_{m}=1$ and (\ref{T11S21}) reduces
to the expression given in \cite{Bord09}. In the same limit, (\ref{T00S22})
reduces to the corresponding result in \cite{Bord09} up to a finite
renormalization term $-m/(96a^{2})$. Note that without this term the
renormalized energy density does not vanish in the limit $m\rightarrow
\infty $. For both minimally and conformally coupled fields the energy
density (\ref{T00S22}) is negative and the stress (\ref{T11S21}) is
positive. For a massless conformally coupled field the VEV $\langle
T_{i}^{k}\rangle _{S^{2}}$ vanishes.

The combination of (\ref{T11S2}) and (\ref{T00S21}), leads to the relation%
\begin{equation}
\langle T_{1}^{1}\rangle _{S^{2}}=\partial _{V}E_{S^{2}},  \label{PresEn}
\end{equation}%
where $V=4\pi a^{2}$ is the volume of the space and $E_{S^{2}}=V\langle
T_{0}^{0}\rangle _{S^{2}}$ is the total vacuum energy. This is the standard
thermodynamical relation between the energy and the pressure.

Now let us consider the case $m^{2}a^{2}+2\xi \leqslant 1/4$. In this case
the function $f(x)$, corresponding to (\ref{phi2S21}), has a branch point in
the right half-plane and the formula (\ref{AP}) needs a generalization. As
the starting point we us the Abel-Plana formula in its initial form (see,
for instance, \cite{Bord09,Saha07Rev}) for series $\sum_{l=0}^{\infty }f(l)$%
. Taking $f(x)=g(x)/\sqrt{x^{2}-x_{0}^{2}}$, by the transformation of the
integration contour in the rhs, the following formula can be obtained%
\begin{eqnarray}
\sum_{l=0}^{\infty }\frac{g(l+1/2)}{\sqrt{(l+1/2)^{2}-x_{0}^{2}}}
&=&\int_{x_{0}}^{\infty }dx\,\frac{g(x)}{\sqrt{x^{2}-x_{0}^{2}}}%
+\int_{0}^{x_{0}}dx\,\frac{\tan (\pi x)}{\sqrt{x_{0}^{2}-x^{2}}}g(x)  \notag
\\
&&-\int_{0}^{\infty }dx\,\frac{g(ix)+g(-ix)}{\sqrt{x^{2}+x_{0}^{2}}(e^{2\pi
x}+1)},  \label{AP2}
\end{eqnarray}%
for $x_{0}<1/2$ and for a function $g(x)$ analytic in the right half-plane.
Note that for the series in (\ref{phi2S21}) one has $g(x)=x$ and the last
integral in (\ref{AP2}) vanishes.

Further evaluation of the renormalized VEVs is similar to that we have
described above for the case $m^{2}a^{2}+2\xi \geqslant 1/4$. Applying to
the series in (\ref{phi2S21}) the formula (\ref{AP2}) with $%
x_{0}^{2}=-\omega _{m}^{2}\geqslant 0$, after the renormalization, for the
VEV of the field squared one gets%
\begin{equation}
\langle \varphi ^{2}\rangle _{S^{2}}=\frac{m}{4\pi }\left[ 1+\frac{\xi -1/6}{%
m^{2}a^{2}}+\frac{|\omega _{m}|}{ma}\int_{0}^{1}dy\,\tan (\pi |\omega _{m}|%
\sqrt{1-y^{2}})\right] .  \label{phi2renb}
\end{equation}%
The VEV of the energy density is found from (\ref{T00S21}). This leads to
the expression%
\begin{equation}
\langle T_{0}^{0}\rangle _{S^{2}}=\frac{m^{3}}{4\pi }\left[ \frac{1}{3}+\,%
\frac{\xi -1/6}{m^{2}a^{2}}-\frac{|\omega _{m}|^{3}}{m^{3}a^{3}}%
\int_{0}^{1}dy\,y^{2}\tan (\pi |\omega _{m}|\sqrt{1-y^{2}})\right] .
\label{T00S22b}
\end{equation}%
The integration constant is determined from the continuity of the energy
density at $\omega _{m}=0$. The stresses are obtained by making use of the
relation (\ref{T11S2}).

\end{document}